\documentclass[useAMS,usenatbib]{mn2e}
\usepackage{graphicx,amssym}
\newcommand{\bc}{\begin{center}}
\newcommand{\ec}{\end{center}}
\newcommand{\wdl}{{\small WDL08 }}
\newcommand{\mor}{{\small MORGANA }}
\newcommand{\hop}{{\small HOP09 }}

\title[Bulge formation in a $\Lambda$CDM cosmology]
      {Times, environments and channels of bulge formation in a $\Lambda$CDM
        cosmology}  
\author[G.~De Lucia et al.]
       {Gabriella De Lucia$^{1}$\thanks{Email: delucia@oats.inaf.it}, 
        Fabio Fontanot$^{1}$, David Wilman$^{2}$, Pierluigi Monaco$^{3,1}$
        \\  
        $^1$INAF - Astronomical Observatory of Trieste, via G.B. Tiepolo 11, 
        I-34143 Trieste, Italy\\
        $^2$Max-Planck-Institut f\"ur Extraterrestrische Physik, 
        Giessenbachstra\ss e, D-85748 Garching, Germany\\
        $^3$Dipartimento di Astronomia, Universit\`a di Trieste, via
        G.B. Tiepolo 11, I-34131 Trieste,Italy} 
\begin{document}

\pagerange{\pageref{firstpage}--\pageref{lastpage}} 
\pubyear{2010}

\maketitle

\label{firstpage}

\begin{abstract}
We analyze predictions from two independently developed galaxy formation models
to study the mechanisms, environments, and characteristic times of bulge
formation in a $\Lambda$CDM cosmogony. For each model, we test different
prescriptions for bulge formation in order to quantify the relative importance
of different channels. Our results show that the strong correlation between
galaxy and halo mass for central galaxies, and the richer merger history of
more massive systems naturally give rise to a strong correlation between galaxy
mass and morphology, and between halo mass and morphological type of central
galaxies. Large fractions of the bulge mass are acquired through major and
minor mergers, but disc instability plays an important role, particularly for
intermediate mass galaxies. We find that the modelling of disc instability
events, as well as of the galaxy merger times, can affect significantly the
timing of bulge formation, and the relative importance of different
channels. Bulge dominated galaxies acquire their morphology through major
mergers, but this can be modified by cooling of gas from the surrounding hot
halo. We find that disc regrowth is a non negligible component of the evolution
of bulge dominated galaxies, particularly for low to intermediate masses, and
at high redshifts.
\end{abstract}


\section{Introduction}
\label{sec:intro}

Perhaps the clearest and most convenient definition of a {\it bulge} is that of
a centrally concentrated stellar distribution, with a smooth and spherical
appearance. Indeed, such a definition underlies the classification scheme
introduced by \citet{Hubble_1926}. In the local Universe, about $60$ per cent
of the total stellar mass of massive galaxies is contained in ellipticals and
bulges \citep{Gadotti_2009}. It is clear then, that understanding how bulges
form and evolve is integral to the question of understanding galaxy formation
and evolution.

Until the early 1980s, bulges were thought to belong to the same family as
elliptical galaxies, and to have formed through the same physical
process(es). Several lines of evidence, however, indicated later that the class
`bulges' represents a heterogeneous family including systems with very
different properties, and likely very different formation and evolutionary
histories. Indeed, significant differences were found between the kinematics of
ellipticals and bulges \citep[e.g.][and references
  therein]{Dressler_Sandage_1983,Davies_Illingworth_1983}. It was also noted
that many bulges exhibit a `boxy' or `peanut' shaped structure at small radii.
This shape, that is unlikely to be due to the gravitational influence of the
disc, was found to be associated with differential cylindrical rotation
\citep{Kormendy_Illingworth_1982}. In the past decades, substantial evidence
has accumulated that many bulges have `disc-like' exponential profiles
\citep[e.g.][]{Andreakis_Sanders_1994,Carollo_etal_2001,Balcells_etal_2003,Fisher_Drory_2008}
and, in some cases, `disc-like' cold kinematics \citep[][and references
  therein]{Kormendy_1993,Pinkney_etal_2003,Kormendy_Kennicutt_2004}. These
systems are now often referred to as `pseudo-bulges', as opposed to `classical
bulges' that are relatively featureless both dynamically and photometrically,
and appear to have a close affinity with elliptical galaxies. 

The current view is that classical bulges are formed through rapid collapse or
hierarchical mergers of smaller objects, and corresponding dissipative gas
processes. Early numerical simulations showed that close interactions can lead
to a strong internal dynamical response, driving the formation of spiral arms
and, in some cases, of strong bar modes. The axisymmetry of these structures
induces a compression of the gas that can fuel nuclear starbursts and/or
nuclear AGN activity \citep[see e.g.][and references
  therein]{Mihos_2004}. Simulations have also shown that the merger of two
spiral galaxies of comparable mass can produce a remnant with structural and
photometric properties resembling those of elliptical galaxies
\citep*[e.g.][]{Toomre_Toomre_1972,Mihos_2004,Springel_DiMatteo_Hernquist_2005}.
On the other hand, pseudo-bulges are thought to originate from the evolution of
disc instabilities such as bars. Early simulations by \citet{Hohl_1971} showed
that bar formation is accompanied by a rearrangement of disc material, which
results in the formation of a high-density central core. Later and more
detailed simulations have confirmed that gravitational instabilities such as
spirals and bars are able to build `bulge-like' structures, either through
vertical resonances or through bending (`buckling') of the bars
\citep[e.g.][]{Combes_etal_1990,Raha_etal_1991,Debattista_etal_2006}.

All these processes are at play in the general framework of hierarchical galaxy
formation: galaxies are supposed to form through the condensation of gas at the
centre of dark matter haloes. Conservation of angular momentum leads to the
formation of a rotationally supported disc. If the cooling is `rapid' (at high
redshift and in relatively small haloes), the short dynamical times lead to an
intense star-burst activity. Mergers and instabilities form `bulges', that can
eventually grow a new disc, provided the system is fed by an appreciable
cooling flow. In this framework then, bulge-dominated galaxies can be
`transitory' systems.  The importance of disc regrowth and its correlation with
the physical properties and/or environment of galaxies has, however, not been
analyzed in detail.

Accurate studies of the structural and physical properties of bulges and
ellipticals are now being carried out \citep[e.g.][at low
  redshift]{Gadotti_2009}. These studies and their extension to higher
redshift, will likely provide important constraints on how the different
population of bulges evolved as a function of cosmic time. It is therefore
interesting to analyze in more detail predictions from recently published
galaxy formation models, with particular regard to the question of what is the
relative role of different physical mechanisms (e.g. mergers vs disc
instability) in the formation of galaxy bulges, and their evolution as a
function of redshift, environment and galaxy mass.

In this paper, we analyze predictions from two independently developed
semi-analytic models of galaxy formation. While the models used in this study
do not allow a fine classification into `bulges', `pseudo-bulges', or `bars' to
be made, they allow us to quantify the amount of mass that is contributed to
the spheroidal components by different `channels' (minor and major mergers, and
disc instability), and to study when and in which environment(s) these
processes take place. Using two different models and, within them, different
prescriptions for the formation of bulges, we are able to analyze how the
relative importance of different channels varies as a function of different
specific physical assumptions. Some of these issues have been addressed using
similar classes of models in previous studies
\citep*{Parry_Eke_Frenk_2009,Benson_Devereux_2010}, and we will comment on
these results below. In this study, we focus on theoretical predictions, and
defer a detailed comparison between model results and observational data to a
future work.

The layout of the paper is as follows. In Section~\ref{sec:simsam}, we
introduce the galaxy formation models used in this study, focusing on those
aspects of the models that are relevant for bulge formation. In
Section~\ref{sec:trends}, we discuss the basic trends predicted as a function
of the galaxy stellar mass and of the parent halo virial mass. In
Sections~\ref{sec:hwhen} and \ref{sec:hwhere}, we analyze the times and
environments that characterize the formation of galaxy bulges through different
channels. In Section~\ref{sec:discless}, we study the formation history of
`elliptical' galaxies and address the issue of disc regrowth. Finally, we
discuss our results, and give our conclusions in Section~\ref{sec:discconcl}.

\section{The galaxy formation models}
\label{sec:simsam}

In this paper, we consider predictions from two different and independently
developed semi-analytic models of galaxy formation within a $\Lambda$CDM
cosmogony. In particular, we use (i) the recent implementation of the Munich
model by \citet{DeLucia_and_Blaizot_2007}, with its generalization to the
{\small WMAP3} cosmology discussed in \citet{Wang_etal_2008}; and (ii) the \mor
model presented in \citet*{Monaco_Fontanot_Taffoni_2007}, and adapted to a
{\small WMAP3} cosmology as described in \citet{LoFaro_etal_2009}.

Comparisons between different specific predictions from these two models have
been discussed in \citet{Fontanot_etal_2009} and \citet{Fontanot_etal_2010}. A
detailed analysis of the prescriptions adopted to model gas cooling and galaxy
mergers is given in \citet{DeLucia_etal_2010}. We refer to the original papers
for more details on the modelling of various physical processes. In the
following, we highlight the main differences between the implementations of
these ingredients, focusing on those physical processes that are relevant for
bulge formation. We also summarize the prescriptions proposed by
\citeauthor{Hopkins_etal_2009a} (2009a), that we have implemented in the two
models used in this study.

\subsection{The \wdl model}
\label{sec:wdl}

{\it Cosmological framework: } The model discussed in \citet[][\wdl
  hereafter]{Wang_etal_2008} takes advantage of N-body simulations that follow
the evolution of $N=540^3$ particles within a comoving box of size $125\,
h^{-1}$Mpc on a side. This corresponds to a particle mass of $7.78\times
10^{8}\,h^{-1}{\rm M}_{\odot}$. In this study, we use their simulation with
{\small WMAP3} cosmological parameters: $\Omega_{\rm m}=0.226$, $\Omega_{\rm
  b}=0.04$, $\Omega_\Lambda=0.774$, $n=0.947$, and $\sigma_8=0.722$. The Hubble
constant is parametrized as $H_0 = 100\, h\, {\rm km\, s^{-1} Mpc^{-1}}$, and
this particular simulation assumes $h=0.743$.

{\it Merger trees:} Simulation data were stored at 64 output times, that are
approximately logarithmically spaced between z=20 and 1, and approximately
linearly spaced in time thereafter. Group catalogues were constructed using a
standard friends--of--friends (FOF) algorithm, with a linking length of $0.2$
in units of the mean particle separation. Each group was then decomposed into a
set of disjoint substructures using the algorithm {\small SUBFIND}
\citep{Springel_etal_2001}, which iteratively determines the self-bound
subunits within a FOF group.  The most massive of these substructures is often
referred to as the {\it main halo}, while this and all other substructures are
all referred to as {\it subhaloes} or {\it substructures}. Only subhaloes that
retain at least $20$ bound particles after a gravitational unbinding procedure
are considered `genuine' subhaloes, therefore setting the subhalo detection
limit to $2.22\times10^{10}\,{\rm M}_{\odot}$. These catalogues were then used
to construct merger history trees of all gravitationally self-bound
substructures, as explained in detail in \citet[][see also
  \citealt{DeLucia_and_Blaizot_2007}]{Springel_etal_2005}.

{\it Galaxy mergers:} At variance with the other model used in this study, the
\wdl one follows dark matter haloes after they are accreted onto larger
systems. This allows the dynamics of satellite galaxies residing in infalling
haloes to be properly followed, until the parent dark matter substructure is
`destroyed' (i.e. falls below the resolution limit of the simulation) by tidal
truncation and stripping \citep[e.g.][]{DeLucia_etal_2004,Gao_etal_2004}. When
this happens, galaxies are assigned a residual surviving time using the
classical dynamical friction formula\footnote{For a detailed discussion of the
  adopted formulation, and for a comparison with different implementations, see
  \citet{DeLucia_etal_2010}.}. The residual merging time is estimated from the
relative orbit of the two merging objects, at the time of subhalo disruption.

In the case of a `minor' merger, the stellar mass of the merged galaxy is
transferred to the bulge component of the remnant galaxy, and a fraction of the
combined cold gas from both galaxies is turned into stars as a result of the
merger. The efficiency of the merger-driven starburst is parametrized
  adopting the formulation proposed by \citet{Somerville_Primack_Faber_2001}: 
\begin{displaymath}
  e_{\rm burst} = \beta_{\rm burst} (m_2/m_1)^{\alpha_{\rm burst}}
\end{displaymath}
where $m_2/m_1$ is the baryonic (gas + stars) mass ratio, and $\alpha_{\rm
  burst} = 0.7$ and $\beta_{\rm burst} = 0.56$ have been chosen to provide a
good fit to the numerical simulations of \citet{Cox_etal_2008}.

All stars that form during the burst, as well as all remaining cold gas, are
added to the disc of the remnant galaxy. If the baryonic mass ratio of the
merging galaxies is larger than $0.3$, we assume that we witness a `major'
merger, that gives rise to a more significant starburst and destroys the disc
of the central galaxy completely, leaving a purely spheroidal stellar
remnant. The remnant galaxy can grow a new disc later on, provided it is fed by
an appreciable cooling flow.

{\it Disc instability:} bulges can also grow through disc instabilities, that
are assumed to take place when the following condition is verified
\citep*{Efstathiou_etal_1982}:
\begin{equation}
  \frac{V_{\rm disc}}{({\rm G}\,m_{\rm disc}/r_{\rm disc})^{1/2}} \lesssim
  \epsilon_{\rm lim}
\label{eq:di}
\end{equation}
In the above equation, $m_{\rm disc}$, $r_{\rm disc}$, and $V_{\rm disc}$ are
the stellar mass, the radius, and the velocity of the disc, respectively. In
this model, $V_{\rm disc} = V_{\rm max}$, and is computed directly from the
underlying $N$-body simulation; $r_{\rm disc}$ is the half-mass radius of the
disc that, for an exponential disc, is equal to $1.68 \times R_{\rm d}$;
$R_{\rm d}$ is the disc scale length, and is computed following
\citet*{Mo_Mao_White_1998}. The model assumes $\epsilon_{\rm lim} = 0.75$, that
is chosen in order to reproduce the observed morphological mix in the local
Universe. For each galaxy, and at each time-step, we check whether the
instability condition is verified and, when this is the case, we transfer
enough stellar mass from the disc to the bulge so as to restore stability.

\subsection{The \mor model}
\label{sec:mor}

{\it Cosmological framework:} The results from the \mor model presented in this
study have been obtained using a $144\, h^{-1}$Mpc box with $N=1000^3$
particles, and adopting a cosmology with $\Omega_{\rm m}=0.24$,
$\Omega_\Lambda=0.76$, $\sigma_8=0.8$, $n=0.96$, and $h=0.72$. The dark matter
data used by \mor are obtained using the code {\small PINOCCHIO}
\citep{Monaco_etal_2002}. This algorithm, based on Lagrangian perturbation
theory, has been shown to provide mass assembly histories of dark matter haloes
that are in excellent agreement with results from numerical simulations
\citep{Li_etal_2007}.

{\it Merger trees:} For details on the construction of merger trees, we refer
to \citet{Monaco_Fontanot_Taffoni_2007} and
\citet{Taffoni_Monaco_Theuns_2002}. We note that {\small PINOCCHIO} does not
provide information on dark matter substructures, so \mor is essentially based
on the equivalent of FOF merger trees.

{\it Galaxy mergers:} In order to model the orbital decay of dark matter
subhaloes and galaxy mergers, \mor uses a slightly updated version of the
fitting formulae provided by \citet{Taffoni_etal_2003}. These take into account
dynamical friction, mass loss by tidal stripping, tidal disruption of
substructures, and tidal shocks. In practice, whenever two (FOF) haloes merge,
the galaxy associated with the smaller halo is assigned a galaxy merger time by
interpolating between the two extreme cases of a `live satellite' (where the
object is subject to significant mass losses) and that of a `rigid' satellite
(that does not suffer a significant mass loss). We refer to the original paper
for details on the implementation.

As in \wdl, \mor distinguishes between minor and major galaxy mergers, using
the same baryonic mass ratio threshold ($0.3$). During a minor merger, the
stellar mass and the cold gas of the accreted satellite are added to the bulge
component of the remnant galaxy, whose disc is unaffected by the merger. During
major mergers, the stellar and gaseous disc of the remnant galaxy are destroyed
and relaxed into a single spheroidal component. The cold gas associated with
the bulge can be efficiently converted into stars, and this occurs on very
short time-scales (effectively triggering a `starburst') during major
mergers. As in \wdl, the remnant galaxy can grow a new disc, out of the gas
cooling at later times.

{\it Disc instability:} For this process, \mor adopts the same stability
criterion as in the \wdl model, but uses different definitions for the mass,
radius and velocity of the disc, and assumes $\epsilon_{\rm lim} = 0.7$
\citep{LoFaro_etal_2009}. As for \wdl, this is chosen in order to
  reproduce the observed morphological mix in the local Universe. In this
model, $m_{\rm disc}$ is the total baryonic mass of the disc, $r_{\rm disc}$ is
the disc scale-length (also computed following \citealt{Mo_Mao_White_1998}),
and $V_{\rm disc}$ is the rotational velocity of the disc, computed as detailed
in \citet{Monaco_Fontanot_Taffoni_2007}. When the instability condition is
verified, half of the baryonic mass of the disc is transferred to the bulge
component. As explained above, the presence of a significant amount of cold gas
in the bulge can trigger a burst of star formation.

{\it Additional processes:} \mor includes additional physical mechanisms that
influence the assembly history of bulges. In particular, the model allows
infall of gas onto an existing bulge, by a fraction equal to the fraction of
disc mass embedded in the bulge. In addition, the model takes into account
tidal stripping of stars in satellites, and assumes that a fraction
($f_{sca}=0.7$ in the standard model) of satellite stars are unbound during
major mergers and incorporated into a `diffuse' light component. In the
following, we will neglect the second process (i.e. we will assume
$f_{sca}=0$). We have verified, however, that this does not alter significantly
any of the results discussed in this study.

\subsection{Model differences}
\label{sec:diff}

The previous sections clarify that the two models used in this study differ in
a number of details. As we will see in the following, some of these are
reflected in significant differences between model predictions. In this
section, we briefly comment on these expectations.

In \citet{DeLucia_etal_2010}, we compared the merger model adopted in \mor with
that employed in \wdl, and showed that the former provides merger times that
are systematically shorter (by an order of magnitude) than those predicted by
the latter. This will likely translate into a different relative importance of
the merger channel. In order to quantify the significance of this different
treatment of galaxy mergers, we will also show results from \mor obtained using
longer merger times.

Another significant difference between the two models is related to the adopted
treatment of disc instability: although both models are based on the criterion
proposed by \citet{Efstathiou_etal_1982}, they adopt different definitions for
the mass, radius and velocity of the disk, and instability events have rather
different consequences. In \wdl, disc instability is evaluated only for the
stellar component, and when instabilities occur, only the stellar mass
necessary to restore stability is transferred from the disc to the bulge. No
cold gas component is associated with the bulge in this model. In \mor, a
significant fraction (half) of the baryonic mass (both gas and stars) of the
disc is transferred to the bulge. This particular treatment avoids a series of
consecutive instability events, that are instead frequent in the \wdl model, in
particular at high redshift. As we will show below, however, this modelling
translates into a much more prominent role of the disc instability channel in
bulge formation.

We stress that both models adopted for disc instabilities are oversimplified,
and provide a very crude description of the complex phenomenology of bar
formation and evolution. In particular, the \wdl model neglects the possibility
that bar formation produces an inflow of gas towards the centre that could fuel
starburst/AGN activity, and that can eventually lead to bar disruption. The
assumption of $V_{\rm disc} = V_{\rm max}$, as well as the use of only the
stellar mass disk in Eq.~\ref{eq:di} are questionable. On the other hand, \mor
makes more realistic assumptions about the disc circular velocity and includes
the gaseous mass present in the disc in Eq.~\ref{eq:di}. In both models, the
outcome of an instability event is modelled in a rather arbitrary way. We note
that present simulations do not provide clear indications about the fraction of
disk mass that gets re-distributed, and how this depends on the halo/galaxy
properties. In addition, the very criterion adopted to tag a disk as unstable
has been questioned in recent studies \citep{Athanassoula_2008}. As we will
show in the following, disk instability has important consequences on model
predictions, and more work is certainly needed in order to improve this aspect
of our modelling. In order to quantify the importance of this process, in the
following we will also show model predictions obtained when the disc
instability channel is switched off. We will refer to these runs as the {\it
  pure mergers model} runs.

Another difference between the two models used in this study is given by how
gas is treated during mergers. In the \wdl model, the merger triggers a burst
that converts a fraction of the combined gas into stars. These stars are added
to the disc component of the remnant galaxy. In the \mor model, all gas and
stars of the secondary are transferred to the bulge of the remnant galaxy, and
the cold gas associated with the bulge is efficiently converted into stars.

Furthermore, we note that the \wdl model accounts for satellite-satellite
mergers, while \mor only considers mergers between satellites and central
galaxies. Finally, the small differences in the cosmological parameters adopted
in the two models have little impact on model predictions.

\subsection{The Hopkins et al.  prescriptions}
\label{sec:hop}

\citeauthor{Hopkins_etal_2009a} (2009a, \hop hereafter) analyzed a suite of
hydrodynamic merger simulations and derived a `gas-fraction dependent merger
model'. We refer to the original paper for a detailed derivation of the model,
while a summary of the key prescriptions can be found in Appendix A of
\citeauthor{Hopkins_etal_2009b} (2009b). In this model, the fraction of cold
gas that participates in the starburst associated with a merger can be written
as:
\begin{displaymath}
f_{\rm burst} = \frac{m_{\rm burst}}{m_{\rm cold}} = 1 - (1 + r_{\rm
  crit}/R_{\rm d}) \cdot \exp({-r_{\rm crit}/R_{\rm d}})
\end{displaymath}
where
\begin{displaymath}
  \frac{r_{\rm crit}}{R_{\rm d}} = \alpha \cdot (1 - f_{\rm gas}) \cdot
  f_{\rm disc} \cdot F(\theta,\mu) \cdot G(\mu)
\end{displaymath}
and $f_{\rm gas} = m_{\rm cold}/(m_{\rm cold}+m_{\rm{*,disc}})$ is the gas disc
fraction, $f_{\rm disc} = (m_{\rm cold}+m_{\rm{*,disc}})/m_{\rm bar}$ is the
disc mass fraction, $m_{\rm bar}$ is the baryonic mass of the galaxy. ${R_{\rm
    d}}$ is the disc scale-length, and $\theta$ is the inclination of the orbit
relative to the disc. Assuming that, before coalescence, the distance of
pericentric passage is $b=2 \cdot R_{\rm d}$ (typical of cosmological mergers),
one obtains:
\begin{displaymath}
\alpha \, F(\theta,\mu) = \frac{0.5}{1 - 0.42 \, \sqrt{1 + \mu} \, cos\theta} 
\end{displaymath}
where the parameter $\alpha$ subsumes details of the stellar profile shape and
bar driven distortion dynamics during a merger. Finally, $G(\mu)$ contains the
dependence on the merger mass ratio, and has the form:
\begin{displaymath}
G(\mu) = \frac{2 \mu}{1 + \mu}
\end{displaymath}
where $\mu = m_2/m_1$. In the literature, and even in the two papers by Hopkins
et al. mentioned above, there are inconsistencies in the definition of
`mass-ratio'. In the implementation of the `Hopkins' prescriptions used in this
study, we define the mass of interest as the baryonic plus tightly bound
central dark matter. Specifically, we have included the dark matter
contribution in the following way: for each galaxy, we store the virial mass of
the parent halo at the time of accretion (i.e. at the last time the galaxy is
central), and add 10 per cent of this mass to the baryonic mass of the merging
galaxies. We have verified, however, that the inclusion of the dark matter
contribution does not affect significantly the results discussed below. The new
stars formed during the merger driven starburst are added to the bulge
component of the remnant galaxy.

The whole stellar mass of the secondary is added to the spheroidal component of
the remnant galaxy. It is further assumed that a fraction of the primary's
stellar disc is transferred to the bulge component of the remnant galaxy, and
is violently relaxed.  Specifically, the mass of the disc that is `destroyed'
is:
\begin{equation}
m_{\rm{disc,destroyed}} = \mu \, m_{\rm{*,disc}}
\label{eq:destr}
\end{equation}
We note that in the two standard models used in this study, the stellar disc of
the primary is always unaffected during minor mergers and completely destroyed
during major mergers. Therefore, we will monitor separately this model
component when analysing results from the runs adopting these prescriptions.

\citeauthor{Hopkins_etal_2009b} (2009b) have investigated the implications of
the proposed model in a cosmological framework, using both empirical
halo-occupation models and the semi-analytic model presented in
\citet{Somerville_etal_2008}. In particular, they claim that their model leads
to a significant suppression of bulge formation in low-mass galaxies, and that
simulations and models that ignore the gas dependence of merger induced
starbursts, have difficulties in reproducing the strong observed
morphology-mass relation.

We have implemented the prescriptions illustrated above in both models used in
this study, and will re-address these issues below. We note that in the runs
adopting the \hop prescriptions, we have always switched off the disc
instability channel. In addition, in the \wdl model, the runs using these
prescriptions assume that the stars formed during merger-driven starbursts are
added to the bulge component of the remnant galaxy (as in
\citeauthor{Hopkins_etal_2009b} 2009b). All other model details and parameters
have been left unchanged.

\section{Dependency on stellar and halo mass}
\label{sec:trends}

In this section, we discuss the basic trends predicted by the two models used
in this study, and analyze how they are modified by switching off the disc
instability channel or adopting the \hop prescriptions discussed in the
previous section.

\begin{figure*}
\bc
\resizebox{16cm}{!}{\includegraphics[]{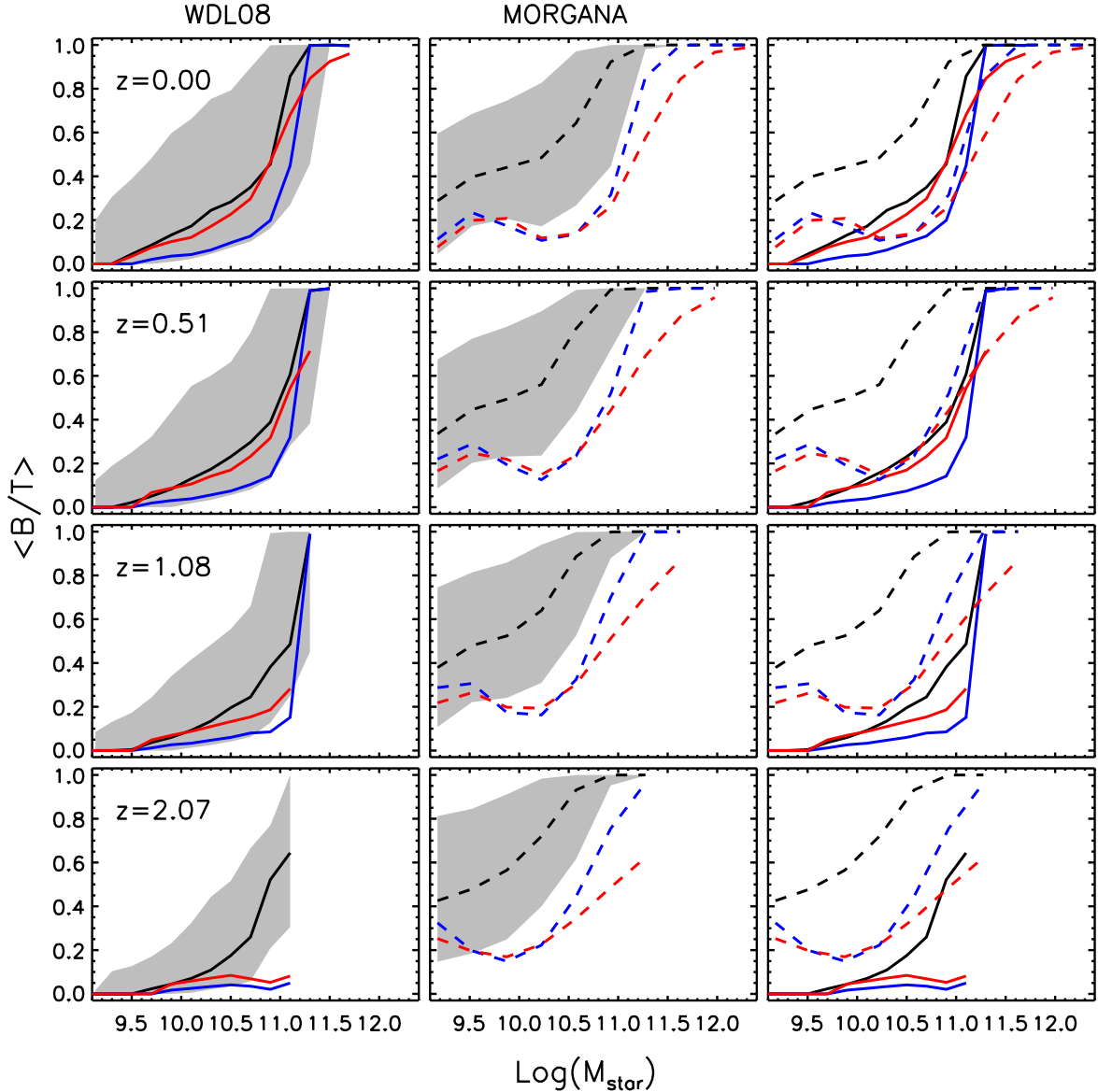}} 
\caption{Median bulge-to-total ratio as a function of the galaxy stellar
  mass. Different rows correspond to different redshifts, while different
  columns correspond to different models: the left column shows results from
  the \wdl model; the middle column shows results from \mor, and the right
  column compares results from the two models. In all panels, black lines
  correspond to the standard models, blue lines correspond to the pure merger
  variant of these models, and red lines show results obtained using the \hop
  prescriptions. The shaded regions in the left and middle columns show the
  15th and 85th percentiles of the distributions obtained for the standard \wdl
  and \mor runs respectively.}
\label{fig:bt_smass}
\ec
\end{figure*}

Fig.~\ref{fig:bt_smass} shows the median (stellar) bulge-to-total ratio as a
function of the galaxy stellar mass, at four different redshifts. The left and
middle columns show predictions from the \wdl and \mor models, respectively,
with different colours used for different physical assumptions. The right
column compares predictions from the two models.

The \wdl model (solid black lines) predicts a strong increase of the
bulge-to-total ratio as a function of the galaxy stellar mass. When the disc
instability channel for bulge formation is switched off (solid blue lines), the
median bulge-to-total ratio decreases for all galaxies but the most massive
ones which are still dominated by the bulge. Compared to this pure mergers run,
the \hop one\footnote{We recall that the disc instability channel has been
  switched off in the run adopting the \hop prescriptions.} predicts a larger
median bulge-to-total ratio for intermediate mass galaxies, but a smaller one
for the most massive galaxies. In the \wdl model, the median bulge-to-total at
fixed stellar mass decreases slightly with increasing redshift. The scatter is
large, as indicated by the shaded regions. This scatter is somewhat reduced
when considering central galaxies only, and it reflects the variation in galaxy
(and halo) merger trees at fixed galaxy stellar mass.

Also the standard \mor model (dashed black lines) predicts an increase of the
median bulge-to-total ratio as a function of stellar mass, but this is somewhat
shallower than that predicted by \wdl. In the pure mergers run (blue dashed
lines), the median bulge-to-total ratio at fixed stellar mass decreases with
respect to the standard run, but it is higher than in the corresponding run of
the \wdl model, particularly at high redshift. The \hop prescriptions provide
predictions that are very close to those of the pure mergers model for
intermediate mass galaxies, but again lower bulge-to-total ratios for the most
massive galaxies. In the standard \mor run, the median bulge-to-total ratio
increases slightly with increasing redshift, but with significant scatter at
fixed stellar mass, as in \wdl.

The reason for the different behaviour obtained for intermediate mass galaxies
when adopting the \wdl version of the \hop prescriptions can be ascribed to the
different treatment of merger driven starbursts in the \wdl model: during minor
mergers, new stars are added to the disc component of the remnant galaxy in the
standard run, while to the bulge component in the \hop run (see
Section~\ref{sec:diff}). In contrast, in both the \hop and the pure mergers run
of the \mor model, newly formed stars are added to the bulge component of the
central galaxy.

Interestingly, the standard \mor run predicts a quite large bulge-to-total
ratio for galaxies of all masses, even at $z\sim 2$, where a very large
fraction of the galaxies have ${\rm B/T} > 0.4$. Even in the pure merger run,
most galaxies have ${\rm B/T} > 0.2$ at this redshift, and the median
bulge-to-total ratio is significantly higher than for \wdl (compare blue dashed
and solid lines in the bottom right panel of Fig.~\ref{fig:bt_smass}). Clearly,
the efficient production of bulges in \mor at high redshift is not simply due
to different assumptions made when the instability criterion is met, and is
likely related to the shorter merger time scales adopted
\citep{DeLucia_etal_2010}. We will come back to this issue later.

\begin{figure*}
\bc
\resizebox{16cm}{!}{\includegraphics[]{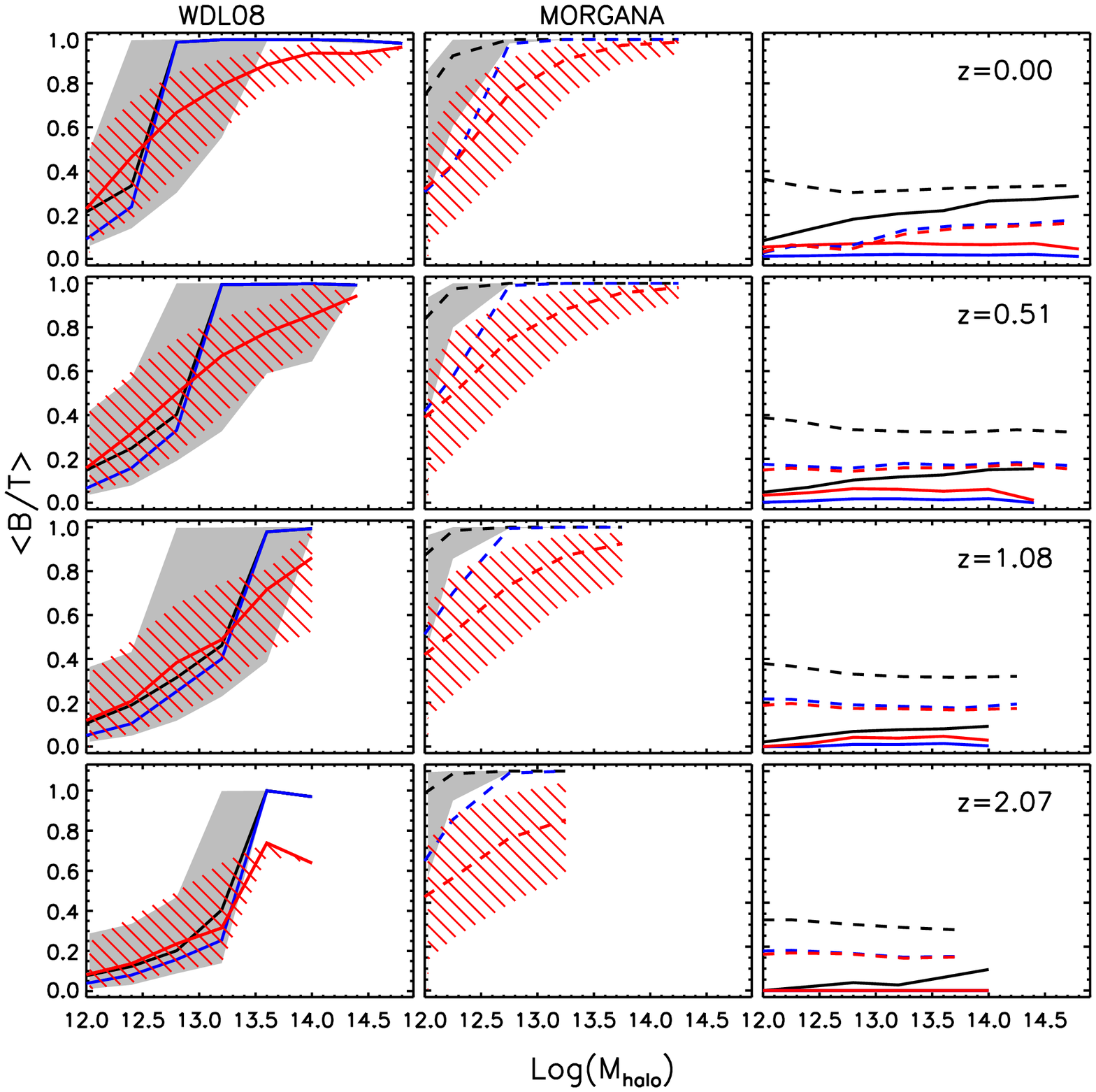}} 
\caption{Median bulge-to-total ratio as a function of the parent halo
  mass. Different rows correspond to different redshifts, while different
  columns and colours correspond to different models, as in
  Fig.~\ref{fig:bt_smass}. The left and middle columns have been obtained for
  central galaxies only, while the right column compares results for satellite
  galaxies. The shaded regions in the left and middle columns show the 15th and
  85th percentiles of the distribution for the standard and \hop runs.}
\label{fig:bt_hmass}
\ec
\end{figure*}

Fig.~\ref{fig:bt_hmass} shows the median bulge-to-total ratio as a function of
the halo mass for the same redshifts as Fig.~\ref{fig:bt_smass}. The left and
middle columns show results for each model only considering central galaxies,
while the right column compares all results for satellite galaxies only.

Both models predict that a large fraction of central galaxies in haloes more
massive than ${\rm log}[{\rm M}_{\rm halo}] \gtrsim 12.5$ are dominated by the
bulge component (the fraction of bulge-dominated central galaxies is
significantly larger for \mor than \wdl). In both models, the \hop
prescriptions provide a somewhat weaker increase of the bulge-to-total ratio as
a function of halo mass, and a reduction of the median bulge-to-total ratio for
the most massive galaxies. The scatter at fixed halo mass is large, reflecting
significant variations in the accretion histories of haloes (and of their
central galaxies) at fixed mass.

The median bulge-to-total ratio of satellite galaxies is relatively low (lower
in the \wdl model than in \mor), and approximately constant as a function of
halo mass. We note that, in these models, once a galaxy is accreted onto a
larger system (i.e. becomes a satellite galaxy) its bulge-to-total ratio is
unaffected, unless it suffers a merger with another satellite
galaxy\footnote{Mergers between satellites are included in \wdl but not in
  \mor.}. In the real Universe, tidal stripping and interactions with other
satellite galaxies (e.g. harassment) are likely to increase the bulge-to-total
ratio of satellite galaxies, increasing the median values plotted in
Fig.~\ref{fig:bt_hmass}. 

Interestingly, almost all central galaxies in haloes with mass slightly larger
than $\sim 10^{12}\, M_{\odot}$ are practically `pure bulges' in the \mor
model, and the vast majority of central galaxies in `Milky-Way type haloes'
(with mass $\sim 10^{12}\,M_{\odot}$) have ${\rm B/T} > 0.6$, at all redshifts
considered. These results suggest that the standard \mor run has difficulties
in forming a Milky-Way like galaxy in haloes of mass similar to that of our
Galaxy \footnote{\citet{Maccio_etal_2010} have studied predictions from \mor
  for the luminosity function of `Milky-Way' satellites. However, in the
  version of the model they use, the disc instability channel is switched
  off.}. This does not appear to be a problem in the standard \wdl run 
\citep[see also][]{DeLucia_Helmi_2008,Li_etal_2010}. We note, however, that
haloes of this mass are only marginally resolved in the simulations used in
this study (with $\sim 1000$ particles in the \wdl simulations).

\section{How and When do bulges form?}
\label{sec:hwhen}

The models we have in hands allow us to ask a number questions about the
formation of bulges: when did bulges form? Was most of their mass assembled
during major or minor mergers? What is the relative importance of disc
instability? How does this vary as a function of redshift? And in which
environments did bulges form?

In order to answer these questions, we have rerun our models and, each time the
mass of the bulge is updated, we have stored the information about the
redshift, the halo mass and the fraction of mass contributed to the final
bulge, distinguishing between major mergers, minor mergers, and disc
instability. For the runs that use the \hop prescriptions, where we do not
include the disc instability channel, we have stored separately the information
about the bulge mass contributed through destruction of the primary's stellar
disc (e.g. Eq.~\ref{eq:destr} in Section~\ref{sec:hop}). We note that our
definition of bulge formation refers to the event adding stars to the bulge of
the selected galaxy or its main progenitor, i.e. the clock is reset for stars
in a secondary galaxy once it merges with a more massive one.

The `channels' defined above correspond generally to a combination of different
physical processes: e.g. in \mor, disc instability triggers both a
re-arrangement of the stellar material originally distributed in the disc, and
the inflow of disc gas towards the centre. This leads, in turn, to the
formation of {\it in situ} new stars - a process that is not included in the
\wdl treatment of disc instability. Analogously, mergers are generally
associated with both a starburst, and a re-arrangement of stars belonging
to the merging galaxies.

\begin{figure*}
\bc
\resizebox{16cm}{!}{\includegraphics[]{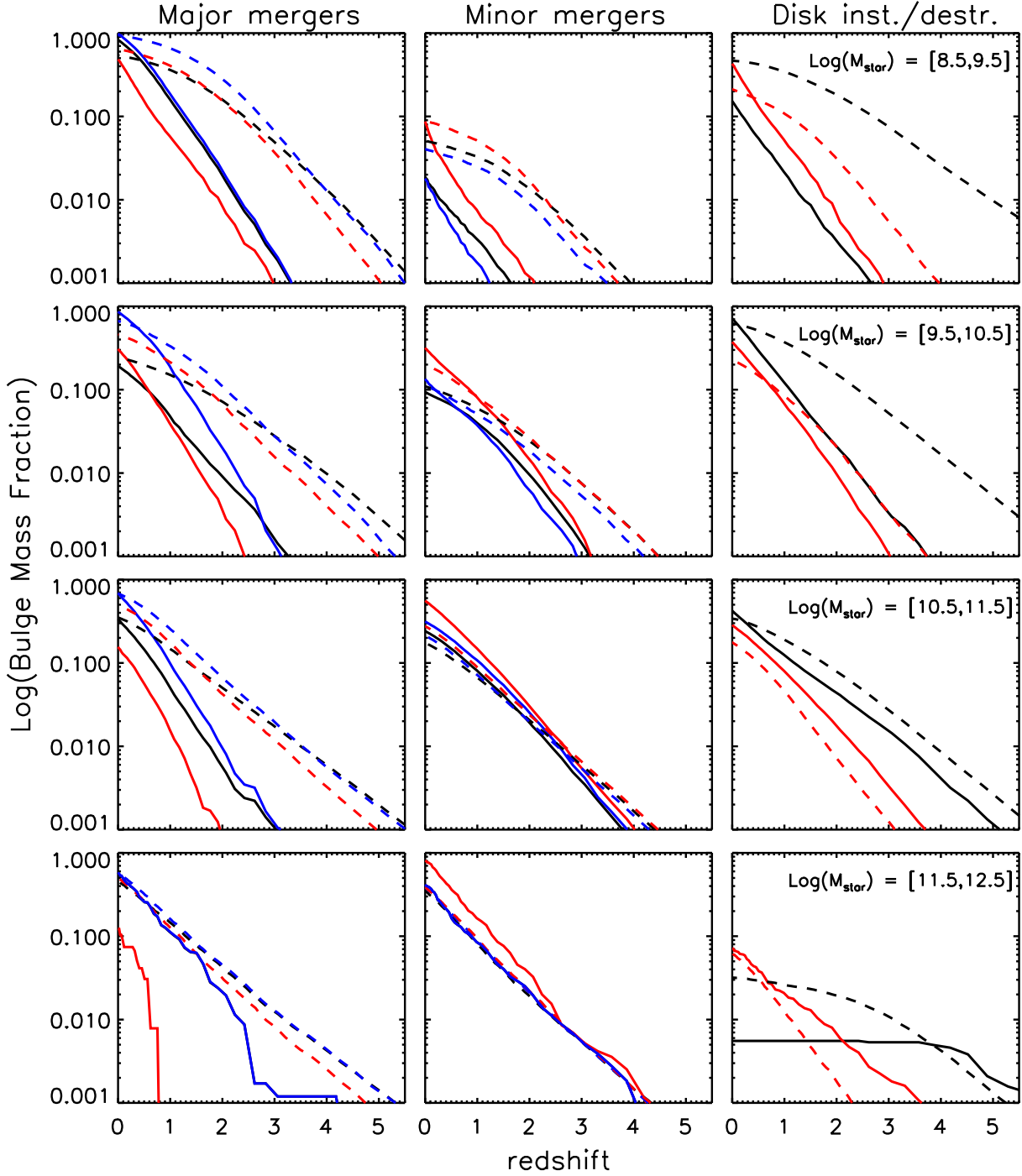}} 
\caption{Fraction of bulge mass formed as a function of redshift through
  different channels (different columns), for galaxies with ${\rm B/T} > 0.4$
  selected at $z=0$. Different rows correspond to different stellar mass bins,
  while different colours and linestyle correspond to results from different
  models as in Fig.~\ref{fig:bt_smass}. Where a black line is not visible, it
  overlaps perfectly with the corresponding blue line. In the right column, the
  black lines indicate the fraction of bulge mass formed through disc
  instability in the standard run, while red lines correspond to the fraction
  of bulge mass formed through destruction of the primary's stellar disc in the
  \hop prescriptions (see Eq.~\ref{eq:destr} in Section \ref{sec:hop}). }
\label{fig:when}
\ec
\end{figure*}

Fig.~\ref{fig:when} shows the fraction of bulge mass contributed through
different channels (different columns) for galaxies selected at $z=0$ as those
having ${\rm B/T}> 0.4$ (qualitatively, the results do not change significantly
when considering all galaxies). We have split model galaxies into four stellar
mass bins, and show the corresponding results in different rows, with the most
massive galaxies shown at the bottom. As in previous figures, solid lines show
results from the \wdl runs, while dashed lines show the corresponding results
from \mor, with different colours referring to different runs. Where
a black line is not visible, it overlaps perfectly with the corresponding blue
line, meaning that switching off disc instability does not affect the
contribution of that particular channel to the final bulge mass (bottom
panels). In the right column, the black lines show the fraction of bulge mass
formed through the disc instability channel in the standard models, while red
lines correspond to the fraction of bulge mass formed through destruction of
the primary's disc in the runs that adopt the \hop prescriptions. We stress
that these two sets of lines have been plotted in the same panels for
convenience, and should not be compared against each other.

Fig.~\ref{fig:when} shows that the contribution from major mergers decreases
with increasing stellar mass, while the contribution from minor mergers
increases. The contribution from disc instability is largest for
intermediate-mass galaxies, and negligible for the most massive galaxies
considered (bottom right panel). Both models and all runs considered share
these trends, with a few notable differences: (i) bulges seem to form earlier
in \mor than in \wdl; (ii) the contribution from disc instability is larger in
the \mor model. For the most massive galaxies, disc instability contributes
less than $\sim 1$ per cent of the final bulge mass, and all instabilities
occurred at high redshift, in the small galaxies that later merged to form
these massive systems. In \mor, the contribution from the disc instability
channel is $\sim 3$ per cent for the most massive galaxies, but it tends to
increase (weakly) since $z\sim 4$. These results can be compared to those that
\citet{Parry_Eke_Frenk_2009} find for the semi-analytic model discussed in
\citet{Bower_etal_2006}. Their Figure 8 shows that instabilities contribute to
the bulges of present day galaxies significantly more than minor and major
mergers, but for the most massive galaxies where the major merger contribution
is dominant. Interestingly, they show that the contribution from disk
instabilities in the `Durham' model is much larger than that in the `Munich'
model (which corresponds to the \wdl model used in this study). The large
contribution from disk instabilities in the Durham model is noted also in
\citet[][see e.g. their Figure 5]{Benson_Devereux_2010}, and is due to the
assumption that instabilities result in the complete collapse of the disc (both
of its stellar and gaseous component - see original paper for
details). Clearly, the different outcome assumed for instability events has
important consequences on the relative importance of different channels to
bulge formation - a conclusion that seems to contradict what found by
\citet[][appendix A2]{Benson_Devereux_2010}. We will come back to this issue
later.

In \mor, the \hop prescriptions result in a contribution from major and minor
mergers that is approximately equal to that found in the standard run. In the
\wdl model instead, the \hop prescriptions result in a systematically lower
contribution from major mergers, and higher contribution from minor mergers, at
all redshifts. This is largely due to the fact that, when adopting the \hop
prescriptions, the \wdl model assumes that the stars formed during all merger
driven starbursts are added to the bulge component of the remnant galaxy
(rather than to the disc in the case of minor mergers, as in the standard
run). The reduced contribution from major mergers relates partly to the
increased efficiency of bulge formation via minor mergers. One notable
consequence of the \hop prescriptions for the \wdl model is for the most
massive galaxies considered: for these, the major mergers channel becomes
important only at $z<1$. Finally, in both models, the disc destruction channel
contributes significantly to the final bulge mass for the intermediate mass
bins considered, but less than $10$ per cent for the most massive galaxies in
the sample.

\begin{figure*}
\bc
\resizebox{16cm}{!}{\includegraphics[]{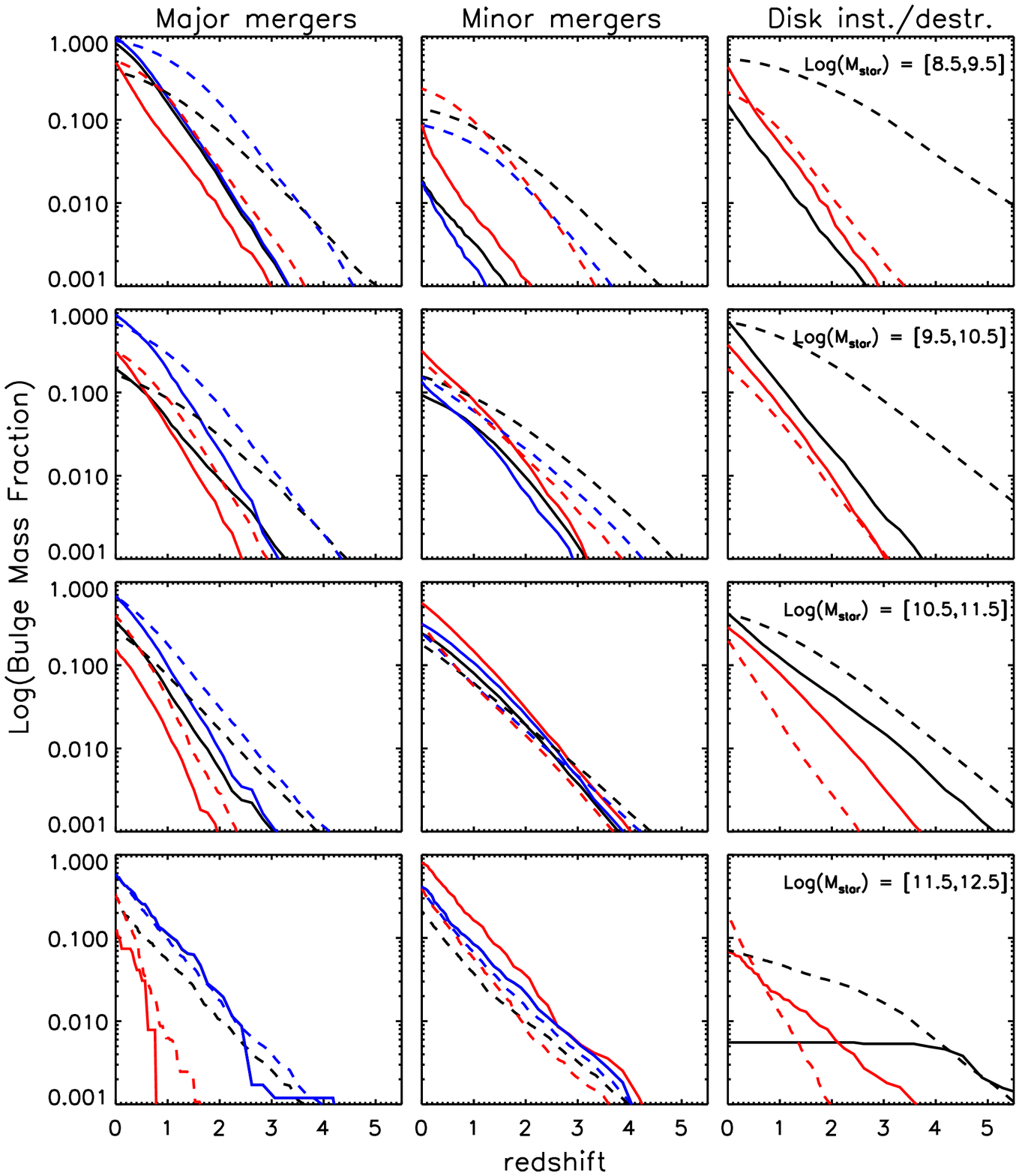}} 
\caption{As in Fig.~\ref{fig:when}, but using longer merger times in the \mor
  runs (see text for details).}
\label{fig:when_longtm}
\ec
\end{figure*}

As mentioned above, the merger model adopted in \mor provides galaxy merger
times that are systematically lower than those used in the \wdl
\citep{DeLucia_etal_2010}. Since a significant fraction of the final bulge mass
is associated with galaxy mergers (both minor and major), systematic
differences between the galaxy merger time-scales are expected to lead to a
systematic difference in the characteristic formation times of galactic
bulges. In order to understand how these differences affect the results
discussed above, we have re-run the \mor models using the same dynamical
friction timescale prescriptions adopted in \wdl.  Results are shown in
Fig.~\ref{fig:when_longtm}. We note that residual merger times are assigned at
the time of halo mergers in \mor, and orbital parameters are re-assigned after
each major merger. In \wdl, residual merger times for satellite galaxies are
instead assigned when the parent dark matter substructures are stripped below
the resolution limit of the simulation. So, although we are using now the same
formulation of dynamical friction in the two models, overall merger times will
still be different.

The figure shows that, when using longer merger times for the \mor runs, bulges
form later, particularly for the most massive galaxies
considered. Interestingly, making merger time-scales longer also increases the
contribution from disc instability in the standard \mor run. This happens
because galaxy discs now have longer times to develop instabilities. Increasing
the merger times in \mor does not account for all differences between the two
models used in this study. In particular, bulge formation still occurs earlier
in \mor than in \wdl for intermediate to low mass galaxies, and disc
instability plays a much more important role in \mor than in the \wdl model,
particularly at high redshift.

\section{How and Where do bulges form?}
\label{sec:hwhere}

In the previous section, we have analyzed {\it when} bulge formation occurs,
and what is the relative importance of different channels at different
times. Another question that can be addressed with our models is: what is the
typical {\it environment} of bulge formation? Does it occur in groups or in the
`field'? And how does the characteristic environment of bulge formation vary as
a function of cosmic time?

\begin{figure*}
\bc
\resizebox{16cm}{!}{\includegraphics[]{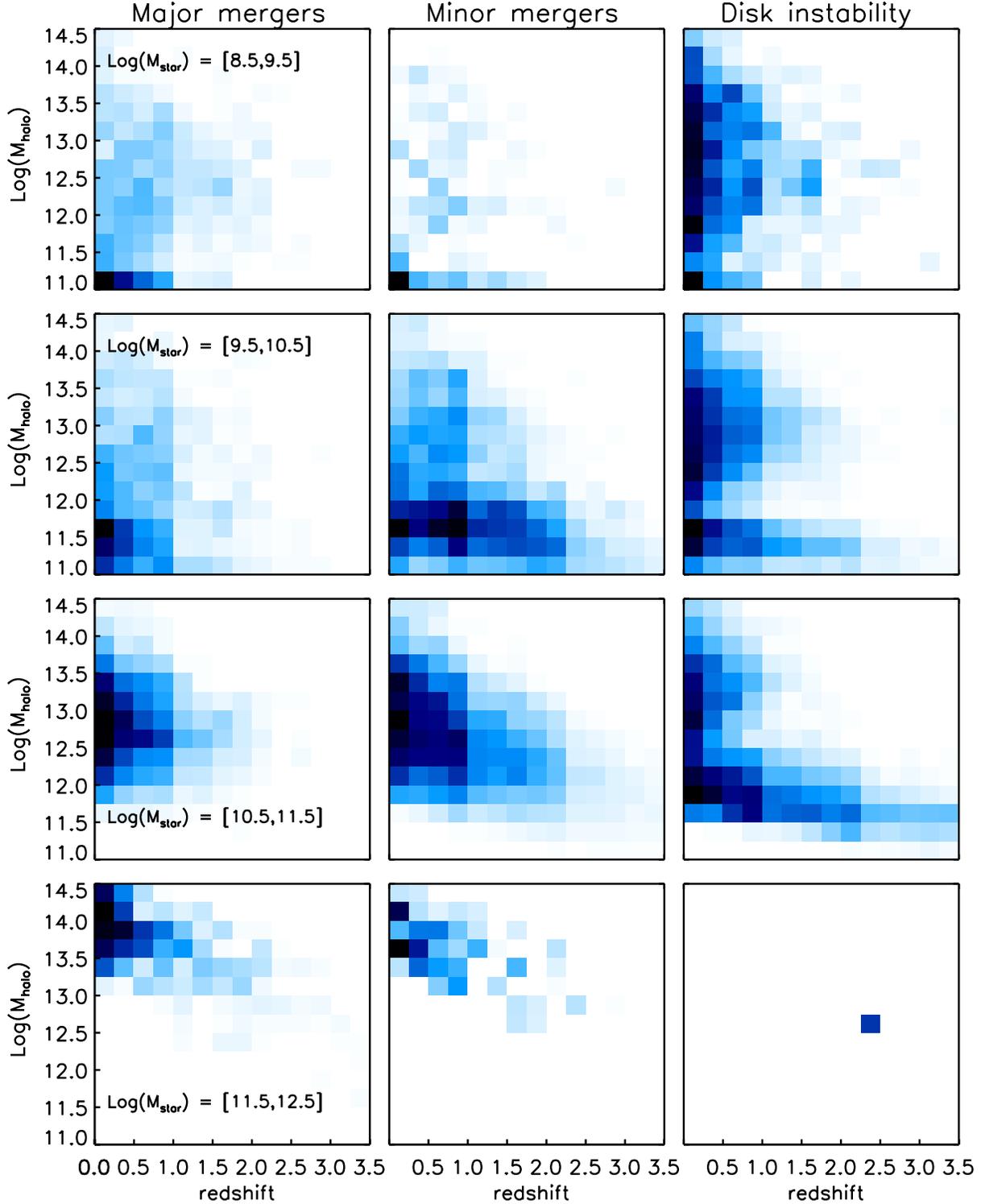}} 
\caption{Fraction of the total bulge mass of galaxies of different stellar mass
  (increasing from top to bottom row) contributed from different channels
  (different columns), as a function of redshift and parent halo mass. For the
  standard \wdl run.}
\label{fig:hesswdl}
\ec
\end{figure*}

\begin{figure*}
\bc
\resizebox{16cm}{!}{\includegraphics[]{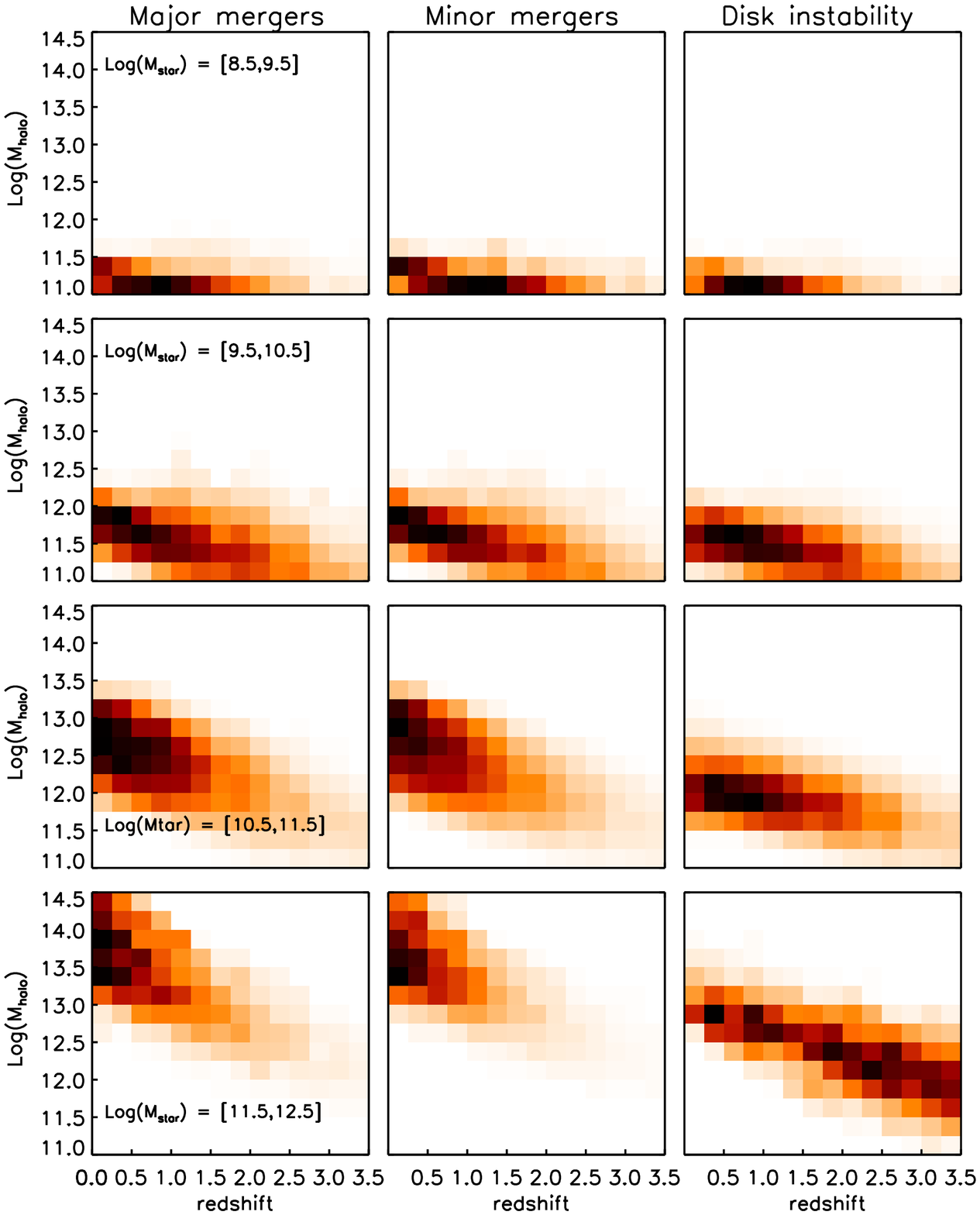}} 
\caption{Same as in Fig.~\ref{fig:hesswdl}, but for the standard \mor run.}
\label{fig:hessmor}
\ec
\end{figure*}

In Figs~\ref{fig:hesswdl} and \ref{fig:hessmor}, we show the fraction of bulge
mass contributed through different channels, as a function of redshift and
parent halo mass, for the \wdl and \mor models respectively. Different rows
correspond to different present day stellar mass bins, while different columns
refer to different channels, as indicated by the legend. Data shown in
Figs~\ref{fig:hesswdl} and \ref{fig:hessmor} have been computed for the
standard model considering all galaxies with ${\rm B/T} > 0.4$, and have been
normalized to the total bulge mass in each mass bin. Therefore, darker regions
in each panel of Figs~\ref{fig:hesswdl} and \ref{fig:hessmor} indicate the
ranges of redshift and halo mass where that particular channel is more
important. Qualitatively, the results shown do not change when including all
galaxies (i.e. without any cut for the bulge-to-total ratio).

The figures show that the typical halo mass where different processes
contribute to bulge formation increases with increasing stellar
mass. Interestingly, for galaxies with ${\rm log}[{\rm M}_{\rm star}] \sim
9-10$, much of the bulge formation occurs in haloes of ${\rm log}[{\rm M}_{\rm
    halo}] < 11.5$ that are only marginally resolved in the simulations used in
this study. One clear difference between the two models is that bulge formation
occurs on a wider range of halo masses (more extended towards larger masses) in
\wdl. This difference is primarily seen for satellite galaxies. Indeed, when
excluding these galaxies, Figs~\ref{fig:hesswdl} and Fig.~\ref{fig:hessmor}
become more similar. For the major and minor mergers channels, this difference
is due to the fact that the \wdl model takes into account mergers between
satellite galaxies, that are not included in \mor. The contribution from
satellite galaxies to the disc instability channel comes from the fact that in
\wdl, the disc radius of a satellite decreases in proportion to the radius of
its dark matter halo. This might not be generally true and could artificially
increase the bulge-to-total ratio of satellite galaxies. We have verified that
fixing the disk radius at the time of
infall\footnote{\citet{Weinmann_etal_2010} have verified that results from the
  model would be virtually unchanged but for the morphology of satellite
  galaxies.} decreases significantly the contribution from disk instability due
to satellite galaxies. We note, however, that our model does not include
physical processes such as tidal stripping and harassment that would again
increase the bulge-to-total ratio of satellites.

Another obvious difference between Figs~\ref{fig:hesswdl} and \ref{fig:hessmor}
is the more pronounced (and extended) contribution from disc instability to the
formation of the most massive bulges in the standard \mor run with respect to
the \wdl model (see also bottom right panel of Fig.~\ref{fig:when}). Finally,
these figures show that there is a somewhat `tighter' correlation between the
mass of the halo and the redshift in \mor. This is likely due to the fact that
mass accretion histories obtained using {\small PINOCCHIO} are `smoother' than
those obtained from numerical simulations.

\section{Ellipticals and disc regrowth}
\label{sec:discless}

In this section, we will focus on galaxies that are dominated by a bulge, and
that we will call `ellipticals' \citep[for a more extended discussion on the
  formation history of elliptical galaxies, see
  also][]{DeLucia_etal_2006}\footnote{Note, however, that a different
  definition of `ellipticals' was adopted in that study.}. More specifically,
we include in the elliptical class all galaxies with at most ten per cent of
the stellar mass in a disc ($B/T > 0.90$). We will address, in particular,
three specific questions: (i) what is the typical stellar mass and environment
of elliptical galaxies? (ii) what is the frequency and relevance of disc
regrowth for these galaxies? (iii) when do galaxies {\it become ellipticals}
and through which physical process(es)?

\begin{figure*}
\bc
\resizebox{15cm}{!}{\includegraphics[]{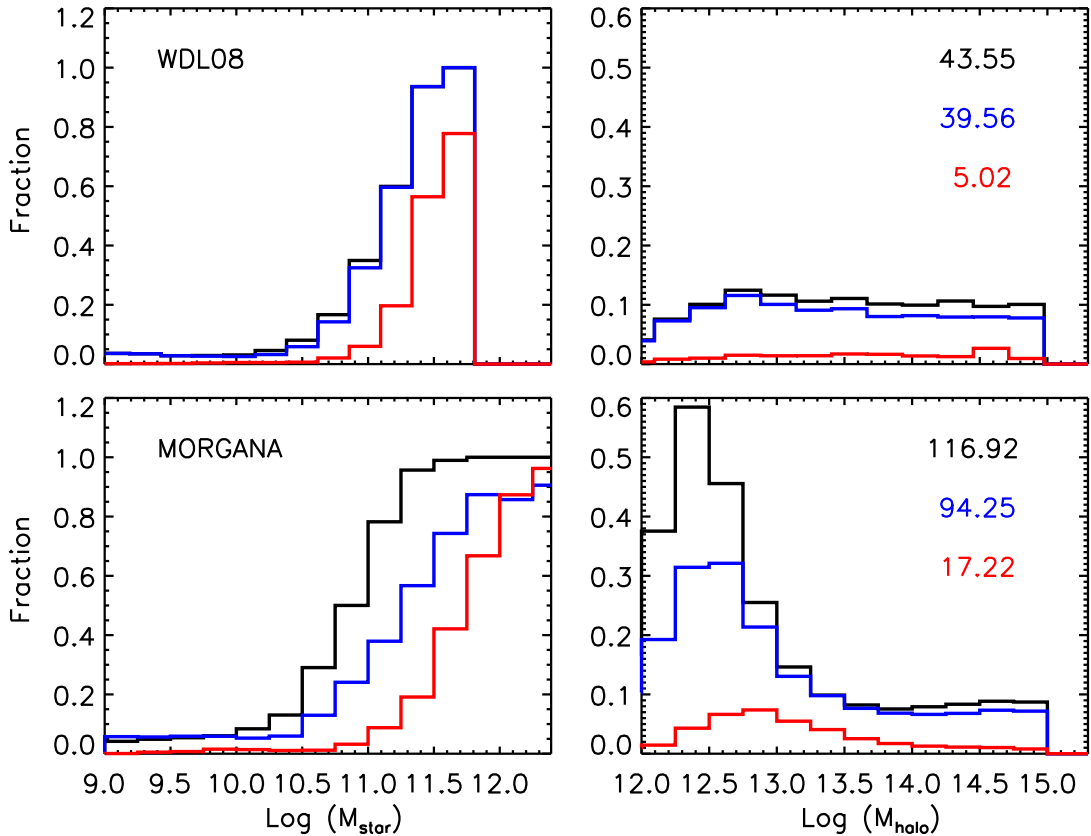}} 
\caption{Distribution of galaxies with ${\rm B/T} > 0.90$ at $z=0$, as a
  function of the galaxy stellar mass (left panels) and of the parent halo mass
  (right panels). The top and bottom rows show results for the \wdl and \mor
  models respectively. Different colours correspond to different runs: black
  lines correspond to the standard models, blue lines correspond to the pure
  mergers variant of these models, and red lines show results obtained using
  the \hop prescriptions. The number densities of galaxies are given in the
  legend of the right panels, and are expressed in units of ${\rm
    Mpc}^{-3}$. The histograms have been normalized dividing by the number of
  galaxies in each stellar or halo mass bin.}
\label{fig:pb}
\ec
\end{figure*}

Fig.~\ref{fig:pb} shows the fraction of galaxies classified as ellipticals, as
a function of the galaxy stellar mass (left panels) and of the parent halo mass
(right panels). Top and bottom panels refer to the \wdl and \mor models,
respectively. As the stellar mass increases, a larger fraction of galaxies are
classified as ellipticals, and the distributions computed from \mor extend to
larger masses than those obtained for \wdl. We note that an excess of massive
galaxies in the \mor model is well documented and is primarily due to an
inefficient suppression of star formation in central galaxies by AGN feedback
\citep[e.g.][]{Fontanot_etal_2009,Kimm_etal_2009}.  Disc instability does not
affect significantly the number and distribution of elliptical galaxies in
\wdl, while the fraction of ellipticals is sensibly reduced when switching off
the disc instability channel in \mor, also at relatively large masses. The
distribution as a function of halo mass is approximately flat for the \wdl
model, and more skewed towards low-mass haloes in \mor (i.e. a larger fraction
of galaxies residing in relatively low-mass haloes are classified as
ellipticals in this model), where a large fraction of ellipticals in relatively
low-mass haloes form through disc instability. The decreasing fraction to
higher halo mass in this model is due to the increasing contribution of
disc-dominated satellite galaxies. Interestingly, the \hop prescriptions tend
to reduce significantly the number of ellipticals in both models. Overall, the
number densities of galaxies that are classified as ellipticals in \mor are
much larger (more than a factor two) than those obtained in \wdl. We note that
model results are in qualitative agreement with observational data indicating
that the total fraction of ellipticals is not expected to be higher in more
massive haloes \citep{Wilman_etal_2009}.

\begin{figure}
\bc
\hspace{-0.7cm}
\resizebox{9cm}{!}{\includegraphics[]{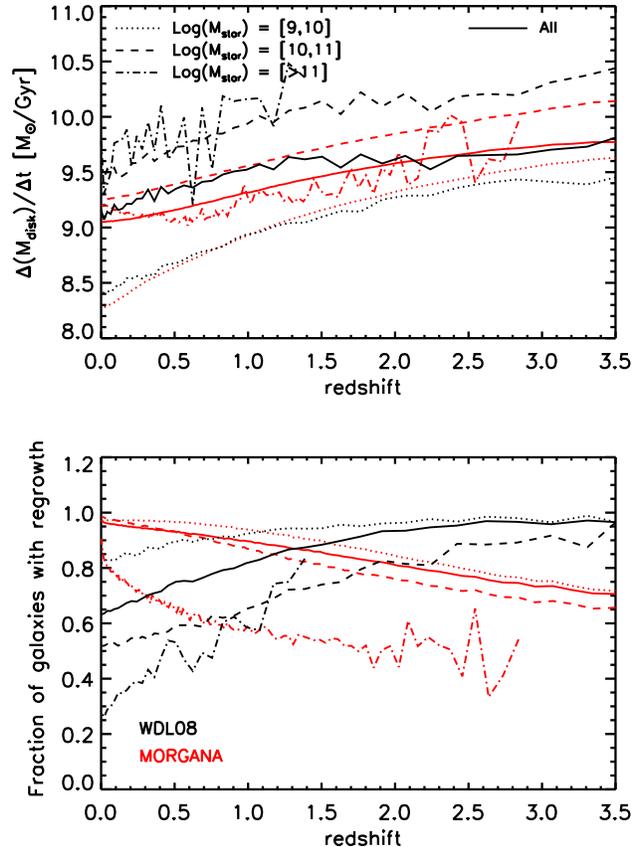}} 
\caption{Top panel: Mean regrowth rate (see text for details). Bottom panel:
  fraction of central galaxies that regrow a disc. Black lines refer to the
  standard \wdl run, while red lines correspond to the standard \mor
  run. Different linestyles correspond to galaxies of different stellar mass,
  as indicated in the legend.}
\label{fig:regrowth}
\ec
\end{figure}

As discussed in Sections~\ref{sec:wdl} and \ref{sec:mor}, bulge dominated
galaxies can regrow a new disc, provided there is enough cooling from the
surrounding hot halo gas. In principle, minor mergers with gas-rich satellites
also form new disc stars in the \wdl model. However, most of the galaxies
accreted onto centrals of relatively massive haloes (where most of the central
ellipticals are) will be gas-poor \citep{DeLucia_and_Blaizot_2007}, and mergers
between satellites are rare. Therefore, the only mechanism through which
galaxies can grow a new disc is by accretion of fresh gas material from the
surrounding hot halo.

In order to quantify the importance of disc regrowth, we have selected all
central elliptical galaxies from our models and analyzed their merger trees,
storing the increase of their stellar disc mass as a function of time, until
they become satellites (or until $z=0$, for central galaxies). The top panel of
Fig.~\ref{fig:regrowth} shows the mean regrowth rate for galaxies in different
stellar mass bins (the stellar mass corresponds to that of the galaxy at the
redshift under consideration). The rates shown in Fig.~\ref{fig:regrowth} have
been normalised to the total number of (central) galaxies experiencing
regrowth. The figure shows that, on average, the rate of disc regrowth
decreases with decreasing redshift, for galaxies of all masses. For the most
massive galaxies considered (dot-dashed lines in Fig.~\ref{fig:regrowth}), the
rate of regrowth averaged over $1\,{\rm Gyr}$ time-scale is always
significantly smaller than the galaxy mass at the corresponding time. This
means that a galaxy that crosses the threshold $B/T=0.90$ when it is already
rather massive, will likely stay above this threshold at any later time. For
lower mass galaxies, the rates are more significant, particularly at high
redshift where galaxies are more gas-rich. On average, however, these galaxies
will not grow a large disc: a galaxy with stellar mass $\sim 10^{10}\,{\rm
  M}_{\odot}$ and $B/T=1$ will regrow a disc containing about 10 per cent of
the mass in $\sim 3-7$~Gyr at $z\sim 0$, or $\sim 1-3$~Gyr at $z\sim 1$. Disc
regrowth is also inhibited by subsequent mergers which increase the
bulge-to-total ratio. These results suggest that disc regrowth is not
significant for the most massive galaxies, but that it represents a non
negligible component in the evolution of low and intermediate mass galaxies,
particularly at high redshift.

The bottom panel of Fig.~\ref{fig:regrowth} shows the fraction of central
galaxies experiencing regrowth in the two models used in this study. At the
highest redshifts considered, the balance between cooling and feedback is such
that more galaxies grow a disc in \wdl than in \mor. As the redshift decreases,
AGN feedback becomes more and more important, particularly for the most massive
galaxies (that are sitting in the most massive haloes). This determines a
significant decrease of the fraction of central galaxies experiencing regrowth
in \wdl: only about $20$ per cent of the most massive galaxies are growing a
new disc, and, as explained above, this occurs at relatively low rates. On the
contrary, the fraction of central galaxies experiencing disc regrowth {\it
  increases} with decreasing redshift in \mor. Albeit the regrowth rates are
low also in this model, almost all galaxies ($\sim 90$ per cent of the most
massive ones) are growing a new disc. This different behaviour can be ascribed
to the different treatment of radio-mode AGN feedback, that is much more
efficient in suppressing cooling in \wdl than in \mor
\citep[e.g.][]{Kimm_etal_2009,Fontanot_etal_2010}.

\begin{figure*}
\bc
\resizebox{16cm}{!}{\includegraphics[]{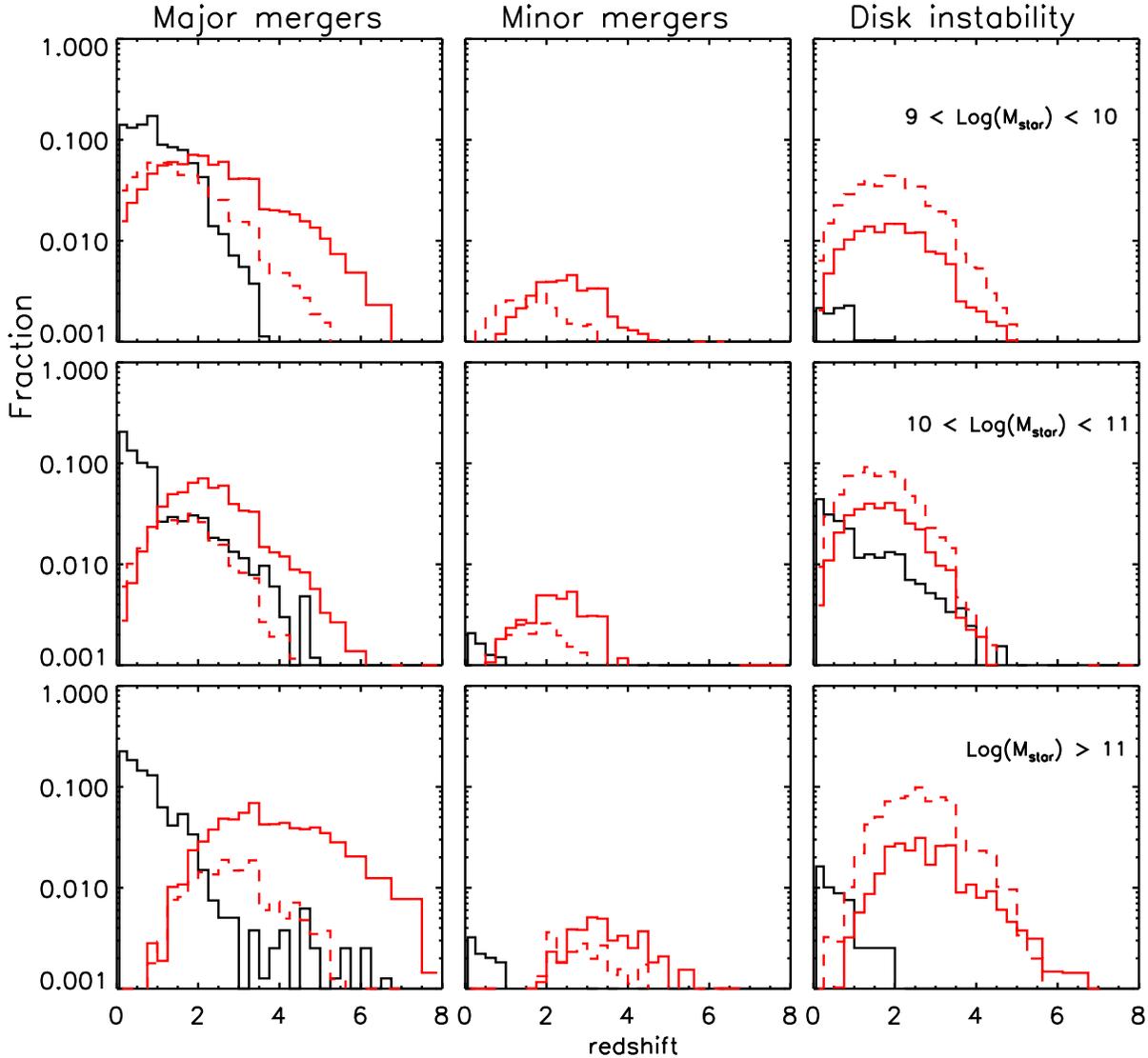}} 
\caption{Distribution of the times when galaxies cross the threshold $B/T=0.9$
  for the first time. Galaxies are split by their final stellar mass
  (different rows) and each column shows the contribution from different
  channels. Galaxy counts are weighted by the fraction of bulge mass
  contributed by each process, and histograms are normalised by the total
  number of galaxies in each mass bin. Black and red lines are used for the
  \wdl and \mor model, respectively. Red dashed lines show results obtained
  from \mor using longer merger times.}
\label{fig:morphz}
\ec
\end{figure*}

Another interesting question to address about elliptical galaxies is when and
through which mechanism(s) they acquired their morphology. We address this
question in Fig.~\ref{fig:morphz} where we show the redshift at which galaxies
cross the threshold $B/T=0.9$ for the first time. Galaxies are split by their
final mass and different columns correspond to different bulge formation
channels. When several processes contribute to make the galaxy cross the
threshold $B/T=0.9$, we have weighted the counts by the fraction of bulge mass
contributed by each channel. The figure shows that most galaxies acquire an
`elliptical' morphology because of a major merger event (left column), while
minor mergers seem to play a negligible role in turning galaxies into
ellipticals, particularly in \wdl. Disc instability is responsible for turning
a relatively large fraction of galaxies into ellipticals in \mor, particularly
at intermediate masses, while it plays a much less prominent role in \wdl. The
distributions obtained from \mor are peaked at redshifts significantly higher
than in \wdl, as a consequence of the significantly shorter merger times and of
the more prominent role of disc instability. Dotted lines in
Fig.~\ref{fig:morphz} show results from \mor obtained adopting longer merger
times (see Section~\ref{sec:hwhen}), and confirms that galaxies become
ellipticals later when longer merger times are adopted. As discussed in
Section~\ref{sec:hwhen}, the figure also shows that this run is characterized
by a larger contribution from disc instability to bulge formation. We stress
that the times plotted in Fig.~\ref{fig:morphz} should not be confused with the
`formation times' of elliptical galaxies, as these times are not related to the
star formation history of these galaxies.

Not all galaxies considered in Fig.~\ref{fig:morphz} are still ellipticals at
$z=0$. As discussed above, however, the regrowth rates are lower for
intermediate-massive galaxies so that a larger fraction of the most massive
galaxies maintain their elliptical morphology down to $z=0$. In particular, we
find that in the \wdl model about 30, 64, and 96 per cent of the galaxies in
each of the mass bins considered (in order of increasing mass) still have $B/T
> 0.9$ at redshift zero. In \mor, the corresponding fractions are 85, 46, and
88 per cent. The fractions increase to about 53, 79, and 98 for the \wdl model
and 89, 61, and 92 per cent for \mor, when considering galaxies with $B/T>0.6$
at $z=0$.

\section{Discussion and conclusions}
\label{sec:discconcl}

In this paper, we have analyzed predictions for the formation of bulges from
two independently developed galaxy formation models. In particular, we have
considered (i) the recent implementation of the Munich model (\wdl) by
\citet{DeLucia_and_Blaizot_2007}, with its generalization to the {\small WMAP3}
cosmology discussed in \citet{Wang_etal_2008}, and (ii) the \mor model
presented in \citet*{Monaco_Fontanot_Taffoni_2007}, and adapted to a {\small
  WMAP3} cosmology as described in \citet{LoFaro_etal_2009}.

The two models include the same channels for bulge formation (mergers and disc
instability), but assume different prescriptions to model these physical
processes. In order to study how results vary as a function of specific
physical assumptions, we have also implemented alternative merger prescriptions
\citep{Hopkins_etal_2009a}, based on results from recent hydrodynamic merger
simulations. In this paper, we have focused on theoretical predictions,
postponing to a forthcoming paper a detailed comparisons with observational
results (Wilman et al., in preparation). In a companion paper (Fontanot et al.,
in preparation), we will study the physical properties and formation histories
of galaxies with no significant bulge component, as predicted by the same
models considered here. 

Both models used in this study, with all different physical assumptions
considered, predict a strong correlation between the galaxy morphology and its
stellar mass, with more massive galaxies having larger bulge-to-total ratios.
For central galaxies, there is also a strong correlation between the morphology
of a galaxy and its parent halo mass, with most of the central galaxies of
haloes with mass larger than $\sim 10^{13}\,{\rm M}_{\odot}$ being dominated by
a bulge. These trends are not surprising, given our assumption that bulges form
during mergers, and the strong correlation between the galaxy mass and the mass
of the parent halo for central galaxies: more massive galaxies will generally
have a richer merger history than their less massive counterparts, and more
massive galaxies will sit at the centre of more massive haloes.

Taking advantage of our models, we have studied in detail the contribution to
bulge formation from different `channels' (major and minor mergers, and disc
instability). Differences arise between the different models and
implementations, but the results at redshift zero can be summarized as follows:

\begin{itemize}
\item[(i)] major mergers dominate the contribution to bulges of galaxies less
  massive than $\sim 10^{10}\, {\rm M}_{\odot}$; 
\item[(ii)] for galaxies more massive than $\sim 10^{10}\, {\rm M}_{\odot}$,
  the contribution from minor and major mergers are comparable;
\item[(iii)] disc instability represents the dominant contribution to the
  formation of bulges of intermediate mass galaxies
  ($\sim~10^{10}-10^{11}\,{\rm M}_{\odot}$).
\end{itemize} 

Qualitatively, our results are in agreement with what found by
\citet{Parry_Eke_Frenk_2009}, although in their model disk instability plays a
more prominent role. It is worrying that such an important contribution to
bulge formation comes from the process that we probably model in the poorest
way (disc instabilities). The results discussed in this paper confirm that
further work is needed in this area in order to improve our galaxy formation
models. This is true not just for the criterion adopted to tag a disc as
unstable (as discussed for example in \citealt{Athanassoula_2008}), but also
for the treatment of these events. In fact, the two models used in this study
assume the same mathematical criterion for disc instability but adopt quite
different assumptions for the physical quantities considered, and model the
outcome of instability events in different ways. In particular, the \wdl model
only transfers to the bulge a fraction of the {\it stellar} disc that is enough
to restore stability. In \mor, half of the {\it baryonic} mass of the disc
(i.e. gas and stars) is transferred to the bulge component each time an
instability episode occurs. At high redshift, where galaxies are more gas rich
and dynamical times are shorter, inflow of gas towards the centre leads to a
rapid and efficient growth of the central spheroidal component in this model.

\begin{figure}
\bc
\resizebox{8.5cm}{!}{\includegraphics[]{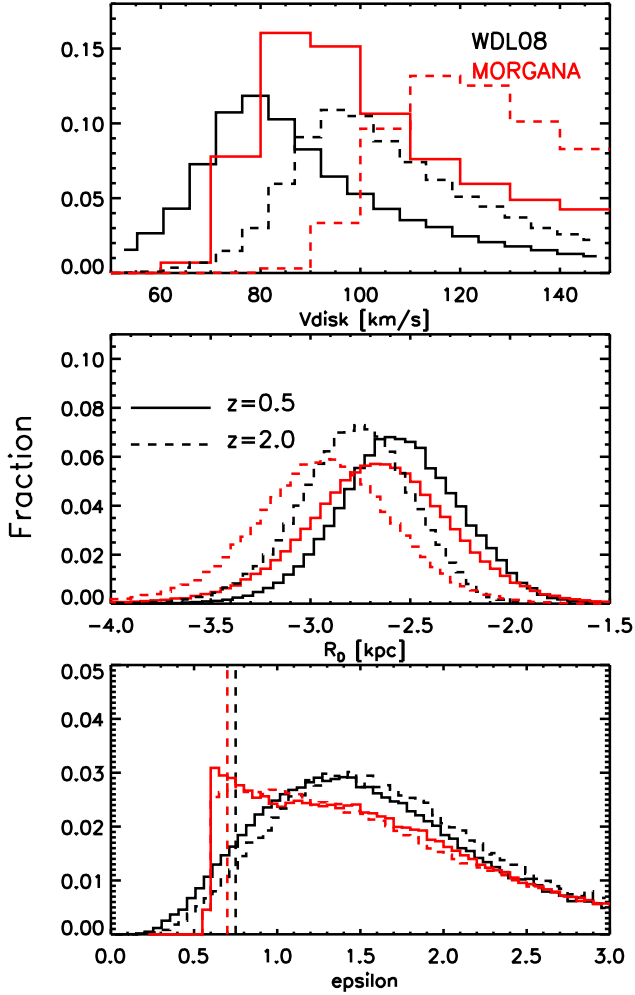}} 
\caption{Distributions of the disc velocity and radius entering Eq.~\ref{eq:di}
  at two different redshifts. Black lines correspond to the standard \wdl
  model, while red lines are for \mor. The bottom panel shows the distribution
  of the left-hand side of Eq.~\ref{eq:di}, and the vertical lines mark the
  threshold for unstable discs adopted in the two models.}
\label{fig:inst}
\ec
\end{figure}

It is instructive to see how different the adopted models are. We can estimate
the fraction of discs that are unstable by looking at the distributions of the
left-hand side of Eq.~\ref{eq:di} for the pure merger models. Results are shown
in the bottom panel of Fig.~\ref{fig:inst}, while the other two panels show the
distributions of the disc velocity and radius entering Eq.~\ref{eq:di} in the
same models. The figure shows that there are systematic differences between
these quantities in the two models. In particular, the disc velocities and disc
scale-lengths used in \mor are systematically larger and smaller than those
adopted in \wdl, respectively. Despite these differences, the fractions of
unstable discs in the two models are comparable: integrating the distributions
shown in the bottom panel up to $\epsilon_{\rm lim} = 0.75$ for \wdl and
$\epsilon_{\rm lim}=0.7$ for \mor (see Section~\ref{sec:simsam}), one obtains
that $\sim 5-7$ per cent of the discs are unstable in both models. This
fraction is approximately independent of redshift. The different treatment of
instabilities, however, leads to a much more prominent role of the disc
instability channel in \mor. To further test our conclusion, we have verified
that the contribution from disc instability (particularly at high redshift) is
significantly reduced in \mor when the fraction of disc mass transferred to the
central spheroidal component is reduced from $0.5$ to $0.1$.

In the framework of our models, bulge dominated galaxies can grow a new disc if
they are fed by appreciable cooling flows. The rates of disc regrowth are
relatively low at low redshift and for massive galaxies. They are, however,
more significant for intermediate and low mass galaxies and at high redshifts,
where ellipticals can change back their morphology and become disc dominated
galaxies at some later time (i.e. the morphology of these galaxies is {\it
  transient}). When an efficient radio-mode feedback is assumed (like in the
\wdl model used here), there is a decline of the typical regrowth rate of the
most massive galaxies at low redshift, and of the fraction of these galaxies
experiencing regrowth. This is expected given that these galaxies live in the
most massive haloes where the radio-mode is assumed to play a major role.

Only a minor fraction of the ellipticals in our models acquire their morphology
through minor mergers. In \wdl, the vast majority of the galaxies become
ellipticals through major mergers, and this occurs at relatively low redshift
($z<2$, only a very small fraction of the galaxies become ellipticals at higher
redshift). In \mor, the change of morphology occurs at higher redshift, and a
large fraction of the galaxies cross the threshold we have adopted to define
galaxies as ellipticals ($B/T=0.9$) as a consequence of disc instabilities. We
have demonstrated that this is due to a combination of significantly shorter
merger times and a different treatment of disc instability events.

The implementation of a gas-fraction dependent merger model provides trends in
bulge-to-total ratio that do not deviate strongly from those of our pure
mergers runs at intermediate masses. This appears to be in contradiction with
the previous claim of \citet{Hopkins_etal_2009b} who find that taking into
account the gas dependence of merger induced starbursts reduces bulge formation
in galaxies less massive than $\sim 10^{10}\,{\rm M}_{\odot}$. We note,
however, that the results from the `simplified model' discussed in Hopkins et
al. are not representative of the results from our semi-analytic models. In
particular, their simplified model provides much more mass in bulges at galaxy
masses lower than $\sim 10^{10}\,M_{\odot}$, where the assumption of a gas
dependent merger model should make the largest difference. We note that our
standard models do include a dependency on the gas available during minor
mergers, and they do so in different ways. The \wdl model assumes that the
stars formed during starbursts associated with these mergers are added to the
disc component of the remnant galaxy, while \mor adds these stars to the bulge
component. The results discussed above show that this model difference does not
significantly influence model results, because minor mergers represent a
relatively minor contribution to bulge formation.

As explained above, our models naturally predict a strong correlation between
the galaxy stellar mass and its morphology. In addition, this correlation
evolves in such a way that {\it fewer} galaxies are bulge dominated at higher
redshift. One of the models used in this study (\mor) does predict an `excess'
of galaxies with large bulge-to-total ratios at intermediate-low masses. As
emphasized above, however, this is not due to the fact that the model neglects
the dependency on gas fraction. Rather, the behaviour of \mor can be ascribed
to galaxy merger times that are significantly shorter than those adopted in
\wdl, and to a different treatment during instability events that leads to an
efficient production of bulges at high redshift. Interestingly, the
implementation of a gas-fraction dependent merger model reduces significantly
the number of bulge dominated galaxies (ellipticals), and delays their
formation. This is a consequence of the assumption that some fraction of the
stellar disc is always preserved, even during major mergers.

It remains to be seen if any of the models discussed in this paper provides a
good agreement with the growing amount of information being accumulated on the
cosmic mophological mix, at different epochs. Such a comparison will certainly
represent an important test-bed for the next generation of models, and provide
stringent constraints on the mechanisms through which bulges are
assembled. While we defer to a forthcoming paper a detailed comparison with
observational data, we note that the predicted bulge-to-total distributions (as
a function of both stellar and halo mass - see Section \ref{sec:trends}), as
well as the total fraction of bulge-dominated galaxies, are affected by the
physical prescriptions adopted to model bulge formation. None of these
observables has been used to `tune' the models in the first place, so that they
can be considered as genuine model predictions and tested against observational
measurements.

We have shown that the contribution to bulge mass from different channels is
also affected by the adopted physical modelling. If `classical bulges' are
primarily formed through mergers and `pseudo-bulges' can be associated with
secular evolution, the results discussed in this paper provide predictions for
the relative importance of these two populations, at different cosmic epochs
and in different environments. Distinguishing pseudo-bulges from classical
bulges is difficult, requires good photometric data and, ideally, also
high-resolution spectroscopic information \citep{Debattista_etal_2005}. Some
statistics are, however, available at low redshift \citep{Gadotti_2009} and can
be used to constrain our models.

\section*{Acknowledgements}
We are grateful to Jie Wang for letting us use the outputs of his
simulations. GDL acknowledges financial support from the European Research
Council under the European Community's Seventh Framework Programme
(FP7/2007-2013)/ERC grant agreement n. 202781. FF acknowledges the support of
an INAF-OATs fellowship granted on `Basic Research' funds. DW acknowledges the
support of the Max-Planck Gesellschaft.

\bsp

\label{lastpage}

\bibliographystyle{mn2e}
\bibliography{bulge_formation}

\begin{thebibliography}{}

\bibitem[\protect\citeauthoryear{{Andredakis} \& {Sanders}}{{Andredakis} \&
  {Sanders}}{1994}]{Andreakis_Sanders_1994}
{Andredakis} Y.~C.,  {Sanders} R.~H.,  1994, \mnras, 267, 283

\bibitem[\protect\citeauthoryear{{Athanassoula}}{{Athanassoula}}{2008}]{Athana%
ssoula_2008}
{Athanassoula} E.,  2008, \mnras, 390, L69

\bibitem[\protect\citeauthoryear{{Balcells}, {Graham},
  {Dom{\'{\i}}nguez-Palmero} \& {Peletier}}{{Balcells}
  et~al.}{2003}]{Balcells_etal_2003}
{Balcells} M.,  {Graham} A.~W.,  {Dom{\'{\i}}nguez-Palmero} L.,    {Peletier}
  R.~F.,  2003, \apjl, 582, L79

\bibitem[\protect\citeauthoryear{{Benson} \& {Devereux}}{{Benson} \&
  {Devereux}}{2010}]{Benson_Devereux_2010}
{Benson} A.~J.,  {Devereux} N.,  2010, \mnras, 402, 2321

\bibitem[\protect\citeauthoryear{{Bower}, {Benson}, {Malbon}, {Helly}, {Frenk},
  {Baugh}, {Cole} \& {Lacey}}{{Bower} et~al.}{2006}]{Bower_etal_2006}
{Bower} R.~G.,  {Benson} A.~J.,  {Malbon} R.,  {Helly} J.~C.,  {Frenk} C.~S.,
  {Baugh} C.~M.,  {Cole} S.,    {Lacey} C.~G.,  2006, \mnras, 370, 645

\bibitem[\protect\citeauthoryear{{Carollo}, {Stiavelli}, {de Zeeuw}, {Seigar}
  \& {Dejonghe}}{{Carollo} et~al.}{2001}]{Carollo_etal_2001}
{Carollo} C.~M.,  {Stiavelli} M.,  {de Zeeuw} P.~T.,  {Seigar} M.,
  {Dejonghe} H.,  2001, \apj, 546, 216

\bibitem[\protect\citeauthoryear{{Combes}, {Debbasch}, {Friedli} \&
  {Pfenniger}}{{Combes} et~al.}{1990}]{Combes_etal_1990}
{Combes} F.,  {Debbasch} F.,  {Friedli} D.,    {Pfenniger} D.,  1990, \aap,
  233, 82

\bibitem[\protect\citeauthoryear{{Cox}, {Jonsson}, {Somerville}, {Primack} \&
  {Dekel}}{{Cox} et~al.}{2008}]{Cox_etal_2008}
{Cox} T.~J.,  {Jonsson} P.,  {Somerville} R.~S.,  {Primack} J.~R.,    {Dekel}
  A.,  2008, \mnras, 384, 386

\bibitem[\protect\citeauthoryear{{Davies} \& {Illingworth}}{{Davies} \&
  {Illingworth}}{1983}]{Davies_Illingworth_1983}
{Davies} R.~L.,  {Illingworth} G.,  1983, \apj, 266, 516

\bibitem[\protect\citeauthoryear{{De Lucia} \& {Blaizot}}{{De Lucia} \&
  {Blaizot}}{2007}]{DeLucia_and_Blaizot_2007}
{De Lucia} G.,  {Blaizot} J.,  2007, \mnras, 375, 2

\bibitem[\protect\citeauthoryear{{De Lucia}, {Boylan-Kolchin}, {Benson},
  {Fontanot} \& {Monaco}}{{De Lucia} et~al.}{2010}]{DeLucia_etal_2010}
{De Lucia} G.,  {Boylan-Kolchin} M.,  {Benson} A.~J.,  {Fontanot} F.,
  {Monaco} P.,  2010, \mnras, 406, 1533

\bibitem[\protect\citeauthoryear{{De Lucia} \& {Helmi}}{{De Lucia} \&
  {Helmi}}{2008}]{DeLucia_Helmi_2008}
{De Lucia} G.,  {Helmi} A.,  2008, \mnras, 391, 14

\bibitem[\protect\citeauthoryear{{De Lucia}, {Kauffmann}, {Springel}, {White},
  {Lanzoni}, {Stoehr}, {Tormen} \& {Yoshida}}{{De Lucia}
  et~al.}{2004}]{DeLucia_etal_2004}
{De Lucia} G.,  {Kauffmann} G.,  {Springel} V.,  {White} S.~D.~M.,  {Lanzoni}
  B.,  {Stoehr} F.,  {Tormen} G.,    {Yoshida} N.,  2004, \mnras, 348, 333

\bibitem[\protect\citeauthoryear{{De Lucia}, {Springel}, {White}, {Croton} \&
  {Kauffmann}}{{De Lucia} et~al.}{2006}]{DeLucia_etal_2006}
{De Lucia} G.,  {Springel} V.,  {White} S.~D.~M.,  {Croton} D.,    {Kauffmann}
  G.,  2006, \mnras, 366, 499

\bibitem[\protect\citeauthoryear{{Debattista}, {Carollo}, {Mayer} \&
  {Moore}}{{Debattista} et~al.}{2005}]{Debattista_etal_2005}
{Debattista} V.~P.,  {Carollo} C.~M.,  {Mayer} L.,    {Moore} B.,  2005, \apj,
  628, 678

\bibitem[\protect\citeauthoryear{{Debattista}, {Mayer}, {Carollo}, {Moore},
  {Wadsley} \& {Quinn}}{{Debattista} et~al.}{2006}]{Debattista_etal_2006}
{Debattista} V.~P.,  {Mayer} L.,  {Carollo} C.~M.,  {Moore} B.,  {Wadsley} J.,
    {Quinn} T.,  2006, \apj, 645, 209

\bibitem[\protect\citeauthoryear{{Dressler} \& {Sandage}}{{Dressler} \&
  {Sandage}}{1983}]{Dressler_Sandage_1983}
{Dressler} A.,  {Sandage} A.,  1983, \apj, 265, 664

\bibitem[\protect\citeauthoryear{{Efstathiou}, {Lake} \&
  {Negroponte}}{{Efstathiou} et~al.}{1982}]{Efstathiou_etal_1982}
{Efstathiou} G.,  {Lake} G.,    {Negroponte} J.,  1982, \mnras, 199, 1069

\bibitem[\protect\citeauthoryear{{Fisher} \& {Drory}}{{Fisher} \&
  {Drory}}{2008}]{Fisher_Drory_2008}
{Fisher} D.~B.,  {Drory} N.,  2008, \aj, 136, 773

\bibitem[\protect\citeauthoryear{{Fontanot}, {De Lucia}, {Monaco}, {Somerville}
  \& {Santini}}{{Fontanot} et~al.}{2009}]{Fontanot_etal_2009}
{Fontanot} F.,  {De Lucia} G.,  {Monaco} P.,  {Somerville} R.~S.,    {Santini}
  P.,  2009, \mnras, 397, 1776

\bibitem[\protect\citeauthoryear{{Fontanot}, {Pasquali}, {De Lucia}, {van den
  Bosch}, {Somerville} \& {Kang}}{{Fontanot} et~al.}{2010}]{Fontanot_etal_2010}
{Fontanot} F.,  {Pasquali} A.,  {De Lucia} G.,  {van den Bosch} F.~C.,
  {Somerville} R.~S.,    {Kang} X.,  2010, ArXiv:1006.5717

\bibitem[\protect\citeauthoryear{{Gadotti}}{{Gadotti}}{2009}]{Gadotti_2009}
{Gadotti} D.~A.,  2009, \mnras, 393, 1531

\bibitem[\protect\citeauthoryear{{Gao}, {White}, {Jenkins}, {Stoehr} \&
  {Springel}}{{Gao} et~al.}{2004}]{Gao_etal_2004}
{Gao} L.,  {White} S.~D.~M.,  {Jenkins} A.,  {Stoehr} F.,    {Springel} V.,
  2004, \mnras, 355, 819

\bibitem[\protect\citeauthoryear{{Hohl}}{{Hohl}}{1971}]{Hohl_1971}
{Hohl} F.,  1971, \apj, 168, 343

\bibitem[\protect\citeauthoryear{{Hopkins}, {Cox}, {Younger} \&
  {Hernquist}}{{Hopkins} et~al.}{2009}]{Hopkins_etal_2009a}
{Hopkins} P.~F.,  {Cox} T.~J.,  {Younger} J.~D.,    {Hernquist} L.,  2009,
  \apj, 691, 1168

\bibitem[\protect\citeauthoryear{{Hopkins}, {Somerville}, {Cox}, {Hernquist},
  {Jogee}, {Kere{\v s}}, {Ma}, {Robertson} \& {Stewart}}{{Hopkins}
  et~al.}{2009}]{Hopkins_etal_2009b}
{Hopkins} P.~F.,  {Somerville} R.~S.,  {Cox} T.~J.,  {Hernquist} L.,  {Jogee}
  S.,  {Kere{\v s}} D.,  {Ma} C.,  {Robertson} B.,    {Stewart} K.,  2009,
  \mnras, 397, 802

\bibitem[\protect\citeauthoryear{{Hubble}}{{Hubble}}{1926}]{Hubble_1926}
{Hubble} E.~P.,  1926, \apj, 64, 321

\bibitem[\protect\citeauthoryear{{Kimm}, {Somerville}, {Yi}, {van den Bosch},
  {Salim}, {Fontanot}, {Monaco}, {Mo}, {Pasquali}, {Rich} \& {Yang}}{{Kimm}
  et~al.}{2009}]{Kimm_etal_2009}
{Kimm} T.,  {Somerville} R.~S.,  {Yi} S.~K.,  {van den Bosch} F.~C.,  {Salim}
  S.,  {Fontanot} F.,  {Monaco} P.,  {Mo} H.,  {Pasquali} A.,  {Rich} R.~M.,
  {Yang} X.,  2009, \mnras, 394, 1131

\bibitem[\protect\citeauthoryear{{Kormendy}}{{Kormendy}}{1993}]{Kormendy_1993}
{Kormendy} J.,  1993, in {H.~Dejonghe \& H.~J.~Habing} ed., Galactic Bulges
  Vol.~153 of IAU Symposium, {Kinematics of extragalactic bulges: evidence that
  some bulges are really disks}.
p.~209

\bibitem[\protect\citeauthoryear{{Kormendy} \& {Illingworth}}{{Kormendy} \&
  {Illingworth}}{1982}]{Kormendy_Illingworth_1982}
{Kormendy} J.,  {Illingworth} G.,  1982, \apj, 256, 460

\bibitem[\protect\citeauthoryear{{Kormendy} \& {Kennicutt} Jr.}{{Kormendy} \&
  {Kennicutt}}{2004}]{Kormendy_Kennicutt_2004}
{Kormendy} J.,  {Kennicutt} Jr. R.~C.,  2004, \araa, 42, 603

\bibitem[\protect\citeauthoryear{{Li}, {De Lucia} \& {Helmi}}{{Li}
  et~al.}{2010}]{Li_etal_2010}
{Li} Y.,  {De Lucia} G.,    {Helmi} A.,  2010, \mnras, 401, 2036

\bibitem[\protect\citeauthoryear{{Li}, {Mo}, {van den Bosch} \& {Lin}}{{Li}
  et~al.}{2007}]{Li_etal_2007}
{Li} Y.,  {Mo} H.~J.,  {van den Bosch} F.~C.,    {Lin} W.~P.,  2007, \mnras,
  379, 689

\bibitem[\protect\citeauthoryear{{Lo Faro}, {Monaco}, {Vanzella}, {Fontanot},
  {Silva} \& {Cristiani}}{{Lo Faro} et~al.}{2009}]{LoFaro_etal_2009}
{Lo Faro} B.,  {Monaco} P.,  {Vanzella} E.,  {Fontanot} F.,  {Silva} L.,
  {Cristiani} S.,  2009, \mnras, 399, 827

\bibitem[\protect\citeauthoryear{{Macci{\`o}}, {Kang}, {Fontanot},
  {Somerville}, {Koposov} \& {Monaco}}{{Macci{\`o}}
  et~al.}{2010}]{Maccio_etal_2010}
{Macci{\`o}} A.~V.,  {Kang} X.,  {Fontanot} F.,  {Somerville} R.~S.,  {Koposov}
  S.,    {Monaco} P.,  2010, \mnras, 402, 1995

\bibitem[\protect\citeauthoryear{{Mihos}}{{Mihos}}{2004}]{Mihos_2004}
{Mihos} J.~C.,  2004, Clusters of Galaxies: Probes of Cosmological Structure
  and Galaxy Evolution, p.~277

\bibitem[\protect\citeauthoryear{{Mo}, {Mao} \& {White}}{{Mo}
  et~al.}{1998}]{Mo_Mao_White_1998}
{Mo} H.~J.,  {Mao} S.,    {White} S.~D.~M.,  1998, \mnras, 295, 319

\bibitem[\protect\citeauthoryear{{Monaco}, {Fontanot} \& {Taffoni}}{{Monaco}
  et~al.}{2007}]{Monaco_Fontanot_Taffoni_2007}
{Monaco} P.,  {Fontanot} F.,    {Taffoni} G.,  2007, \mnras, 375, 1189

\bibitem[\protect\citeauthoryear{{Monaco}, {Theuns}, {Taffoni}, {Governato},
  {Quinn} \& {Stadel}}{{Monaco} et~al.}{2002}]{Monaco_etal_2002}
{Monaco} P.,  {Theuns} T.,  {Taffoni} G.,  {Governato} F.,  {Quinn} T.,
  {Stadel} J.,  2002, \apj, 564, 8

\bibitem[\protect\citeauthoryear{{Parry}, {Eke} \& {Frenk}}{{Parry}
  et~al.}{2009}]{Parry_Eke_Frenk_2009}
{Parry} O.~H.,  {Eke} V.~R.,    {Frenk} C.~S.,  2009, \mnras, 396, 1972

\bibitem[\protect\citeauthoryear{{Pinkney}, {Gebhardt}, {Bender}, {Bower},
  {Dressler}, {Faber}, {Filippenko}, {Green}, {Ho}, {Kormendy}, {Lauer},
  {Magorrian}, {Richstone} \& {Tremaine}}{{Pinkney}
  et~al.}{2003}]{Pinkney_etal_2003}
{Pinkney} J.,  {Gebhardt} K.,  {Bender} R.,  {Bower} G.,  {Dressler} A.,
  {Faber} S.~M.,  {Filippenko} A.~V.,  {Green} R.,  {Ho} L.~C.,  {Kormendy} J.,
   {Lauer} T.~R.,  {Magorrian} J.,  {Richstone} D.,    {Tremaine} S.,  2003,
  \apj, 596, 903

\bibitem[\protect\citeauthoryear{{Raha}, {Sellwood}, {James} \& {Kahn}}{{Raha}
  et~al.}{1991}]{Raha_etal_1991}
{Raha} N.,  {Sellwood} J.~A.,  {James} R.~A.,    {Kahn} F.~D.,  1991, \nat,
  352, 411

\bibitem[\protect\citeauthoryear{{Somerville}, {Hopkins}, {Cox}, {Robertson} \&
  {Hernquist}}{{Somerville} et~al.}{2008}]{Somerville_etal_2008}
{Somerville} R.~S.,  {Hopkins} P.~F.,  {Cox} T.~J.,  {Robertson} B.~E.,
  {Hernquist} L.,  2008, \mnras, 391, 481

\bibitem[\protect\citeauthoryear{{Somerville}, {Primack} \&
  {Faber}}{{Somerville} et~al.}{2001}]{Somerville_Primack_Faber_2001}
{Somerville} R.~S.,  {Primack} J.~R.,    {Faber} S.~M.,  2001, \mnras, 320, 504

\bibitem[\protect\citeauthoryear{{Springel}, {Di Matteo} \&
  {Hernquist}}{{Springel} et~al.}{2005}]{Springel_DiMatteo_Hernquist_2005}
{Springel} V.,  {Di Matteo} T.,    {Hernquist} L.,  2005, \mnras, 361, 776

\bibitem[\protect\citeauthoryear{{Springel}, {White}, {Jenkins}, {Frenk},
  {Yoshida}, {Gao}, {Navarro}, {Thacker}, {Croton}, {Helly}, {Peacock}, {Cole},
  {Thomas}, {Couchman}, {Evrard}, {Colberg} \& {Pearce}}{{Springel}
  et~al.}{2005}]{Springel_etal_2005}
{Springel} V.,  {White} S.~D.~M.,  {Jenkins} A.,  {Frenk} C.~S.,  {Yoshida} N.,
   {Gao} L.,  {Navarro} J.,  {Thacker} R.,  {Croton} D.,  {Helly} J.,
  {Peacock} J.~A.,  {Cole} S.,  {Thomas} P.,  {Couchman} H.,  {Evrard} A.,
  {Colberg} J.,    {Pearce} F.,  2005, \nat, 435, 629

\bibitem[\protect\citeauthoryear{{Springel}, {White}, {Tormen} \&
  {Kauffmann}}{{Springel} et~al.}{2001}]{Springel_etal_2001}
{Springel} V.,  {White} S.~D.~M.,  {Tormen} G.,    {Kauffmann} G.,  2001,
  \mnras, 328, 726

\bibitem[\protect\citeauthoryear{{Taffoni}, {Mayer}, {Colpi} \&
  {Governato}}{{Taffoni} et~al.}{2003}]{Taffoni_etal_2003}
{Taffoni} G.,  {Mayer} L.,  {Colpi} M.,    {Governato} F.,  2003, \mnras, 341,
  434

\bibitem[\protect\citeauthoryear{{Taffoni}, {Monaco} \& {Theuns}}{{Taffoni}
  et~al.}{2002}]{Taffoni_Monaco_Theuns_2002}
{Taffoni} G.,  {Monaco} P.,    {Theuns} T.,  2002, \mnras, 333, 623

\bibitem[\protect\citeauthoryear{{Toomre} \& {Toomre}}{{Toomre} \&
  {Toomre}}{1972}]{Toomre_Toomre_1972}
{Toomre} A.,  {Toomre} J.,  1972, \apj, 178, 623

\bibitem[\protect\citeauthoryear{{Wang}, {De Lucia}, {Kitzbichler} \&
  {White}}{{Wang} et~al.}{2008}]{Wang_etal_2008}
{Wang} J.,  {De Lucia} G.,  {Kitzbichler} M.~G.,    {White} S.~D.~M.,  2008,
  \mnras, 384, 1301

\bibitem[\protect\citeauthoryear{{Weinmann}, {Kauffmann}, {von der Linden} \&
  {De Lucia}}{{Weinmann} et~al.}{2010}]{Weinmann_etal_2010}
{Weinmann} S.~M.,  {Kauffmann} G.,  {von der Linden} A.,    {De Lucia} G.,
  2010, \mnras, 406, 2249

\bibitem[\protect\citeauthoryear{{Wilman}, {Oemler}, {Mulchaey}, {McGee},
  {Balogh} \& {Bower}}{{Wilman} et~al.}{2009}]{Wilman_etal_2009}
{Wilman} D.~J.,  {Oemler} A.,  {Mulchaey} J.~S.,  {McGee} S.~L.,  {Balogh}
  M.~L.,    {Bower} R.~G.,  2009, \apj, 692, 298

\end{thebibliography}

\end{document}